\documentclass[11pt,a4paper]{article}
\pdfoutput=1
\usepackage{jheppub}

\usepackage{multirow, graphicx,amssymb,url,mathrsfs,amsmath}
\usepackage{wrapfig,boxedminipage,setspace,subfigure,epsfig}
\usepackage{amsxtra,amstext,latexsym,dsfont,amsfonts}
\usepackage{color,eucal}

      \renewcommand{\b}{\beta}

\newcommand{\dd}{\mathrm{d}}

\newcommand{\abs}[1]{\left| #1 \right|}

\renewcommand{\Im}{{ \rm Im}}

\title{Coherent/incoherent metal transition in a holographic model}

\author[a]{Keun-Young Kim,}
\author[a]{Kyung Kiu Kim,}
\author[b]{Yunseok Seo,}
\author[c]{and Sang-Jin Sin}

\emailAdd{fortoe@gist.ac.kr}
\emailAdd{kimkyungkiu@gmail.com}
\emailAdd{yseo@hanyang.ac.kr}
\emailAdd{sjsin@hanyang.ac.kr}

\affiliation[a]{ School of Physics and Chemistry, Gwangju Institute of Science and Technology,
Gwangju 500-712, Korea
}
\affiliation[b]{Research Institute for Natural Science, Hanyang University, Seoul 133-791, Korea}
\affiliation[c]{ Department of Physics, Hanyang University, Seoul 133-791, Korea }

\abstract{
We study AC electric($\sigma$), thermoelectric($\alpha$), and thermal($\bar{\kappa}$) conductivities in a holographic model, which is based on  3+1 dimensional Einstein-Maxwell-scalar action. There is momentum relaxation due to massless scalar fields linear to spatial coordinate.  The model has three field theory parameters: temperature($T$), chemical potential($\mu$), and effective impurity($\beta$). At low frequencies, if  $\beta < \mu$, all three AC conductivities($\sigma, \alpha, \bar{\kappa}$) exhibit a Drude peak modified by pair creation contribution{(coherent metal)}.  The parameters of this modified Drude peak are obtained analytically. In particular, if $\beta \ll \mu$ the relaxation time of electric conductivity approaches to $2\sqrt{3} \mu/\beta^2$ and the modified Drude peak becomes a standard Drude peak.  If $\beta > \mu$ the shape of peak deviates from the Drude form{(incoherent metal)}.  At intermediate frequencies($T<\omega<\mu$), we have analysed numerical data of three conductivities($\sigma, \alpha, \bar{\kappa}$) for a wide variety of parameters, searching for scaling laws, which are expected from either experimental results on cuprates superconductors or some holographic models.
In the model we study, we find no clear signs of scaling behaviour.}


\arxivnumber{1409.8346}

\begin{document}

\maketitle

\section{Introduction}

Holographic methods (gauge/gravity duality) provide novel tools to study many properties of strongly correlated systems by analysing the corresponding higher dimensional gravity theories~\cite{CasalderreySolana:2011us,Hartnoll:2009sz,Herzog:2009xv, Iqbal:2011ae}.   In particular it gives a new  way of computing transport coefficients such as viscosity, relaxation time, and electric/thermal conductivities as well as various equilibrium thermodynamic quantities.  In this paper we focus on electric, thermoelectric, and thermal conductivities of strongly coupled systems by holographic methods.

The early works on this subject, `holographic conductivity', have dealt with the systems with translation invariance~\cite{Hartnoll:2009sz}. However, any system with finite charge density and translation invariance will exhibit an infinite electric DC conductivity. The reason is straightforward: A constant electric field will accelerate charges indefinitely because there is no momentum dissipation, which is implied by  translation invariance.  Real condensed matter systems will not have translation symmetry. It is broken by a background lattice or impurities.  To remedy this infinite conductivity problem, there have been a number of proposals to introduce the momentum dissipation effect in the framework of holography. They fall into two classes: models with inhomogeneous boundary conditions(IBC) and homogeneous boundary conditions(HBC)\footnote{There is an earlier conceptually different idea. It considers a model of a small number($N_f$) of charged degrees of freedom in a bath of a large number($N_c$) of neutral degrees of freedom. If $N_f$ is parametrically much smaller than $N_c$, the momentum of charged degrees of freedom can be absorbed into a bath.  For example, see \cite{Karch:2007pd, Hartnoll:2009ns, Faulkner:2010zz, Faulkner:2013bna}.}

In IBC models, one gives some bulk fields inhomogeneous boundary conditions breaking translation invariance explicitly \cite{Horowitz:2012ky,Horowitz:2012gs,Horowitz:2013jaa,Ling:2013nxa,Chesler:2013qla,Donos:2014yya}.  One may introduce a spatially modulated scalar field $\phi$ or temporal $U(1)$ gauge field $A_t$ mimicking ionic lattice. In the context of AdS black hole of Einstein-Maxwell-scalar system, we may understand the translation symmetry breaking by the Ward identity \eqref{ward1}
\begin{equation}
\nabla^\nu \langle T_{\nu\mu} \rangle = F_{\mu\nu}\langle J^\nu \rangle +  \partial_\mu \phi  \langle \mathcal{O} \rangle \,,
\end{equation}
where the right hand side may not vanish due to a spatially modulated $\phi$ or $A_t$.

In HBC models, one does not impose explicit inhomogeneous boundary conditions, but find a way to break translation invariance effectively. A few models have been studied.
Massive gravity approach~\cite{Vegh:2013sk,Davison:2013jba,Blake:2013bqa,Blake:2013owa} introduces mass terms for some gravitons. It breaks bulk diffeomorphism invariance and consequently violates the conservation of the stress-energy tensor in the boundary field theory.  Some models exploit a continuous global symmetry of the bulk theory~\cite{Donos:2013eha,Donos:2014uba,Donos:2014yya}, where, for example, the global phase of a complex scalar field breaks translational invariance.\footnote{Some of these models may be related to IBC models~\cite{Donos:2013eha}. In a similar spirit, there are models  utilising a Bianchi VII$_0$ symmetry to construct black holes dual to helical lattices~\cite{Donos:2012js, Donos:2014oha}.}  In \cite{Andrade:2013gsa}, a simple model with massless scalar fields linear in spatial coordinate, breaking translation symmetry, 
was introduced\footnote{This model may be understood also based on \cite{Donos:2013eha}. A single massless complex scalar with constant $\varphi$ in (2.6) of \cite{Donos:2013eha} gives rise to a massless axion linear in the $x_1$ direction.}. It was extended further in \cite{Gouteraux:2014hca, Taylor:2014tka}.

On a technical level, IBC models require to solve complicated coupled partial differential equations(PDE) because of explicit inhomogeneous boundary condition.  An advantage of HBC models is that they allow to deal with coupled ordinary differential equations(ODE) because the stress tensor still remains independent of field theory directions and all bulk fields can be treated as functions of the holographic direction. This technical advantage enables us to analyse a model more easily and extensively. Thus it will make possible more analytic and universal understanding on momentum dissipation mechanism at strong coupling, even  though its microscopic field theory interpretation is unclear yet. 

In this paper, we study AC electric, thermoelectric, and thermal conductivities of a HBC model proposed in  \cite{Andrade:2013gsa}, focusing on a Drude nature at low frequencies and scaling laws at intermediate frequencies. 
The model we study is based on the Einstein-Maxwell-scalar action with negative cosmological constant.  Massless scalar fields($\psi_i$) linear to spatial coordinate are considered so that translation symmetry is broken.  Because they enter the stress tensor through the derivative of scalar fields($\partial_M \psi_i$) the bulk fields such as metric and gauge field still can be homogeneous in field theory direction. Furthermore, to have isotropic bulk fields the identical scalar field is  introduced for every field theory spatial direction. In this model, the DC electric conductivity\cite{Andrade:2013gsa}, thermoelectric and thermal conductivity\cite{Donos:2014cya} were computed analytically and our focus is on AC conductivities. AC electric conductivity was also studied in \cite{Taylor:2014tka} and here we analyse it in greater detail as well as thermoelectric and thermal conductivities. For AC conductivities in other HBC models including massive gravity models we refer to \cite{Amoretti:2014zha,Amoretti:2014mma,Aprile:2014aja,Donos:2014yya}.

At low frequencies, the Drude peak of electric conductivity has been observed in many holographic models with momentum dissipation. For example see \cite{Horowitz:2012ky,Horowitz:2012gs,Horowitz:2013jaa,Ling:2013nxa}.
\begin{equation} \label{DrudeIntro}
\sigma(\omega) = \frac{K \tau}{1 - i \omega \tau} \,,
\end{equation}
where $K$ and $\tau$ were determined numerically. The Drude model was originally derived from a quasi-particle picture. However, it was shown that this Drude-like peak can be realised even when there is no quasi-particle picture at strong coupling if the translation symmetry is broken weakly~\cite{Hartnoll:2012rj}. 
In this context, metal without quasi-particle can be divided into two classes: coherent metal with a Drude peak and incoherent metal without a Drude peak~\cite{Hartnoll:2014lpa}. However, since our model is based on AdS-RN black brane solution, there will be a  term containing the contribution  from pair production affected by net charge density, which we denoted by $\sigma_Q$. This suggests the following modified Drude form
\begin{equation} \label{DrudeIntro2}
\sigma(\omega) = \frac{K \tau}{1 - i \omega \tau} + \sigma_Q \,.
\end{equation}
Since, in our model, there is a parameter $\beta$ (the slope of massless scalar fields in \eqref{groundsol4}) controlling the strength of the translation symmetry breaking, we may investigate how coherent/incoherent metal phase is realised.\footnote{The same question was addressed based on analytic DC conductivities in \cite{Gouteraux:2014hca, Donos:2014oha}.} Indeed, In our model, we find that when 
$\beta < \mu$, the momentum dissipation is Drude like while $\beta > \mu$ it is not.   If the peak is Drude-like 
we obtain analytic expressions for $K,\tau$ and $\sigma_Q$.  For $\beta \ll \mu$, $\sigma_Q$ can be ignored and a modified Drude form is reduced to a standard Drude from. Also we confirm the sum rule is satisfied for both cases, Drude and non-Drude. For thermoelectric and thermal conductivities, qualitatively the same results are obtained.

At intermediate frequencies, $T < \omega < \mu$, where $T$ is temperature and $\mu$ is chemical potential, it was shown experimentally that certain high temperature superconductors in the normal phase exhibit scaling law

\begin{equation} \label{mod1}
\sigma = \frac{B}{\omega^{\gamma}}e^{i\frac{\pi}{2} \gamma} \sim \left( \frac{i}{\omega} \right)^{\gamma}\,,
\end{equation}
where $\gamma \approx 2/3$ and $B$ is constant~\cite{Marel:2003aa}.
This scaling law has been studied also in holographic models in a following modified form. 
\begin{equation} \label{mod2}
\sigma =\left( \frac{B}{\omega^{\gamma}} + C\right)e^{i\frac{\pi}{2} {\tilde{\gamma}}}\,,
\end{equation}
where $\gamma, \tilde{\gamma}, B$ and $C$ are constants to be fitted. In models studied in~\cite{Horowitz:2012ky,Horowitz:2012gs,Ling:2013nxa}  scaling behaviours have been produced while in \cite{Donos:2013eha,Taylor:2014tka,Donos:2014yya} no scaling law has been observed. In our model we have analysed electric, thermoelectric, and thermal conductivities in a wide range of parameters for both scaling laws \eqref{mod1} and \eqref{mod2}.  
However it seems that there is no robust scaling law, which agrees to the conclusion in \cite{Taylor:2014tka}. 

From holographic perspective, the computation of electric, thermoelectric, and thermal conductivities are 
related to the Dynamics of three bulk fields fluctuations(metric, gauge, scalar fields). Their dynamics are determined by 
equations of motion, a system of second order coupled ODEs. From the on-shell quadratic action for these fluctuations we can read off 
the retarded Green's functions relevant to three conductivities. 
In the case that many bulk fields are coupled, the computation of the holographic retarded Green's functions is not very straightforward.  To facilitate solving this important problems
we introduce a systematic numerical method following \cite{Amado:2009ts,Kaminski:2009dh} adapted to our purpose. This method,  used to compute conductivities in this paper, can be applied to other models and problems. 
It will be useful especially when many fields are coupled and the system 
has constraint coming from the residual gauge symmetry. 

This paper is organised as follows.
In section \ref{sec2},  after reviewing Einstein-Maxwell theory with massless
scalar fields in general, we focus on a specific ground state solution to introduce momentum relaxation. 
To set up the stage for AC conductivities, we summarise equations for small fluctuations of relevant metric, gauge and scalar fields around the ground state.
In section \ref{method}, we present a general numerical method to compute retarded Green's functions
when many fields are coupled. By using this method, 
in section \ref{RN}, we compute AC electric, thermoelectric, and thermal conductivities. At low frequencies we focus on the shape of the peak, Drude or non-Drude, and at intermediate frequencies we search for possible scaling laws. In section \ref{Conc} we conclude.

\section{  AdS-RN black branes with scalar sources} \label{sec2}

In this section we briefly review the holographic model of momentum relaxation studied in \cite{Andrade:2013gsa}.  We summarize essential minimum to set up stage for our study, AC conductivities, and refer to \cite{Andrade:2013gsa,Taylor:2014tka} for more details and extensions.

\subsection{General action}
Let us start with the Einstein-Maxwell action on a four dimensional manifold $M$ with boundary $\partial M$
\begin{equation}
S_{\mathrm{EM}} = \int_{M}  \dd^{4}x \sqrt{-g} \left[   R - 2 \Lambda -\frac{1}{4}F^2    \right]
- 2 \int_{\partial M} \dd^3 x   \sqrt{-\gamma} K \,,
\end{equation}
where  $\Lambda = -\frac{3}{ l^2}$  is a negative cosmological constant and $F= \dd A$ is the field strength for a $U(1)$ gauge field $A$.  We have chosen units such that the gravitational constant $16 \pi G$ and the cosmological constant $l$ are equal to $1$ . The second term is the Gibbons-Hawking term required for a well defined variational problem with Dirichlet boundary conditions. $\gamma$ is the determinant of the induced metric $\gamma_{\mu\nu}$ at the boundary and $K$ is the trace of the extrinsic curvature.  In order to have a momentum relaxation effect, we include two free massless scalars
\begin{equation}
S_\psi = \int_M \dd^{4}x \sqrt{-g} \left[  - \frac{1}{2}\sum_{I=1}^{2} (\partial\psi_I)^2  \right] \,.
\end{equation}
The action $S_{\mathrm{EM}} + S_\psi $
implies equations of motion\footnote{Index convention: $M,N,\cdots = 0,1,2,r$, and $\mu,\nu,\cdots = 0,1,2$, and $i,j,\cdots = 1,2$.}
\begin{align}
&R_{MN} = \frac{1}{2}g_{MN} \left( R - 2 \Lambda -\frac{1}{4}F^2  - \frac{1}{2}\sum_{I=1}^{2} (\partial\psi_I)^2 \right) +\frac{1}{2}  \sum_{I}  \partial_M \psi_I \partial_N \psi_I +\frac{1}{2} {F_M}^P F_{NP}
\label{Rmn} \\
&\nabla_M F^{MN} =0 \,,  \qquad
\nabla^2 \psi_I =0 \,. \label{Epsi}
\end{align}
%
%
Given the solutions of these equations of motion, the holographically renormalised action($S_\mathrm{ren}$) \cite{Bianchi:2001kw} is obtained by the on-shell action of
\begin{equation}
S_\mathrm{ren} = S_{\mathrm{EM}}  + S_\psi + S_\mathrm{c} \,,
\end{equation}
where $S_c$ is the counter term
\begin{align}
S_\mathrm{c}=\int_{\partial  M}  \dd x^3 \sqrt{-\gamma} \left( - 4 -R[ \gamma]  + \frac{1}{2}  \sum_{I=1}^2 \gamma^{\mu\nu}  \partial_\mu \psi_I \partial_\nu\psi_I    \right) \,,
\end{align}
which is required to cancel out the divergence from  $S_{\mathrm{EM}} + S_\psi $.

For a general understanding of $S_{\mathrm{ren}}$, it is useful to employ the Fefferman-Graham coordinate system
\begin{equation}
\dd s^2 = \frac{\dd \rho^2}{\rho^2} + \frac{1}{\rho^2} g_{\mu\nu} \dd x^{\mu} \dd x^{\nu} \,,
\end{equation}
where the conformal boundary is at $\rho=0$. For gauge field,  we choose radial gauge $A_\rho = 0$.
Near the boundary the solutions are expanded as
\begin{equation} \label{nearb0}
\begin{split}
&g_{\mu\nu} =   g^{(0)}_{\mu\nu} + \rho^2 g^{(2)}_{\mu\nu} + \rho^3 g_{\mu\nu}^{(3)}+\cdots,  \\
&A_\mu=A_\mu^{(0)} + \rho A_\mu^{(1)}+ \cdots,\\
&\psi_I = \psi_I^{(0)} + \rho^2 \psi^{(2)}_I + \rho^3 \psi^{(3)}_I + \cdots\,,
\end{split}
\end{equation}
where leading terms $g^{(0)}_{\mu\nu}, A_\mu^{(0)}, \psi_I^{(0)} $ are chosen to be functions of the boundary coordinates($x^\mu$), which correspond to the sources of the operators in the dual field theory. The analysis of equations \eqref{Rmn}-\eqref{Epsi} near the boundary gives
some constraints.
First, $ g^{(2)}_{\mu\nu}$ and $\psi^{(2)}_I $ are completely fixed in terms of the leading terms.
$g_{\mu\nu}^{(3)}, A_\mu^{(1)},$ and $\psi^{(3)}_I$ are not fixed but have to satisfy
\begin{equation} \label{constraint1}
\nabla^\mu_{(0)} A_\mu^{(1)} =0\,, \quad \mathrm{Tr} \ g^{(3)}_{\mu\nu} = 0 \,, \quad
\nabla^{\nu}_{(0)} g_{\mu\nu}^{(3)} = \psi_I^{(3)} \partial_\mu \psi_I^{(0)} + \frac{1}{3} F_{\mu\nu}^{(0)} A^{(1)\nu} \,,
\end{equation}
where $\nabla^\mu_{(0)} $ is the covariant derivative with $g_{\mu\nu}^{(0)}$.
To completely determine $g_{\mu\nu}^{(3)}, A_\mu^{(1)},$ and $\psi^{(3)}_I$ in terms of given leading terms, we should solve the equations with an appropriate (incoming) boundary condition at the horizon.

With small fluctuations, the renormalisation on shell action up to linear order in fluctuations reads
\begin{equation}
S_{\mathrm{ren}}^{(1)}=\int_{\partial M} \dd x^3 \sqrt{-g^{(0)}} \left(\frac{3}{2} g^{(3)\mu\nu} \delta g_{\mu\nu}^{(0)}
+ 3\psi_{I}^{(3)} \delta \psi_{I}^{(0)}  + A^{(1)\mu}  \delta A_{\mu}^{(0)}  \right)  \,,
\end{equation}
where the leading terms $\delta g^{(0)}_{\mu\nu},  \delta A_\mu^{(0)}$ and $\delta \psi_I^{(0)} $ are interpreted as sources for dual field theory operators: the stress energy tensor $ T^{\mu\nu}$, a $U(1)$ current $ J^\mu$, and a scalar operator ${O}_{I}$ respectively.  Their expectation values are
\begin{equation} \label{onepoint}
\langle T^{\mu\nu} \rangle =  3 g^{(3)\mu\nu} \,, \quad  \langle J^\mu \rangle  = A^{(1)\mu} \,,
\quad \langle O_I \rangle = 3 \psi^{(3)}_I \,.
\end{equation}
The constraint \eqref{constraint1} in terms of the one point function \eqref{onepoint} yields the Ward identities
\begin{align}
\nabla_\mu \langle J^{\mu}  \rangle &= 0 \,, \qquad
\langle T^\mu_\mu \rangle = 0 \,, \quad \\
\nabla^\nu \langle T_{\mu\nu} \rangle &= \langle  O_I \rangle \nabla_\mu \psi_I^{(0)} + F_{\mu\nu}^{(0)} \langle J^\nu \rangle \,, \label{ward1}
\end{align}
which correspond to the invariance of the renormalised action under a $U(1)$ transformation ($\delta A_{\mu}^{(0)}$), a constant Weyl transformation($\xi^\mu = \delta^\mu_\rho \sigma_\rho $), and the coordinate transformation generated by a vector field $\xi^\mu = \xi^\mu(x^\nu)$, $\xi^\rho = 0$.

\subsection{AdS-RN black brane} \label{22}
We want to study the field theory at finite charge density and finite temperature with momentum dissipation.
A gravity dual will be a charged black brane solution with broken translation symmetry. Indeed
the equations \eqref{Rmn} - \eqref{Epsi} admit the following solutions \cite{Bardoux:2012aw}
\begin{align}
\dd s^2 &= G_{MN} \dd x^{M} \dd x^{N} =  -  f(r) \dd t^2 +  \frac{\dd r^2}{f(r)}  +  r^2 \delta_{ij} \dd x^i \dd x^j \,,  \label{groundsol1}  \\
& \quad f(r)= r^2 - \frac{ \beta^2}{ 2 } - \frac{m_0}{r} + \frac{ \mu^2   }{4} \frac{r_0^{2}}{r^{2}} \,, \qquad
m_0 = r_0^3 \left(  1+\frac{\mu^2}{4 r_0^2} - \frac{\beta^2}{2 r_0^2}     \right)  \label{groundsol2} \\
A&= \mu \left(  1- \frac{r_0}{r}   \right)\dd t    \,,  \label{groundsol3} \\
\psi_I &= \beta_{Ii} x^i = \beta \delta_{Ii} x^i\,, \label{groundsol4}
\end{align}
which is reduced to AdS-Reissner-Nordstrom(AdS-RN) black brane solutions for $\beta=0$. Here we have taken special $\beta_{Ii}$, which satisfies $\frac{1}{2}\sum_{I=1}^2   \vec\beta_{I} \cdot \vec\beta_{I}= \beta^2 $ for general cases\footnote{One can easily obtain the general case by spatial rotation in the $x_1- x_2$ plane}.
These analytic solutions have been reported in \cite{Bardoux:2012aw} and explored further in the context of momentum relaxation in \cite{Andrade:2013gsa}.  Even though two scalar fields($\psi_I$) are spatially dependent functions, metric and gauge field are not, thanks to equal contributions from two scalars for two spatial coordinates.
However, with only one scalar field, the solutions are anisotopic and this case has been studied in \cite{Iizuka:2012wt, Cheng:2014qia}.

The solutions \eqref{groundsol1} - \eqref{groundsol4} are characterised by three parameters: $r_0$, $\mu$,  and $\beta$.  $r_0$ is the black brane horizon position($f(r_0) =0$)
and can be replaced by temperature $T$ for the dual field theory:
\begin{equation}
T = \frac{f'(r_0)}{4\pi} = \frac{1}{4\pi} \left( 3r_0 - \frac{\mu^2+ 2\beta^2}{4r_0}  \right) \,,
\end{equation}
from which, $r_0$ yields
\begin{equation} \label{r0}
r_0 = \frac{2\pi}{3} \left( T+\sqrt{T^2 + 3(\mu/4\pi)^2 + 6(\beta/4\pi)^2}  \right)\,.
\end{equation}%

The parameter $\mu$ is the boundary value of $A_t$ identified with the chemical potential in the dual field theory and
$\mu r_0$ corresponds to the charge density according to \eqref{onepoint}.
$\beta$ is the parameter which controls  momentum relaxation.
The parameter $m_0$ obtained by the condition, $f(r_0) =0$,  is a function of $\mu, T, \beta$ and turns out to be proportional to the energy density. In summary, for solutions \eqref{groundsol1} - \eqref{groundsol4}, one point function \eqref{onepoint} is
\begin{equation} \label{Ttt}
\langle{T^{tt}}\rangle = 2m_0 \,, \quad \langle{T^{xx}}\rangle = \langle{T^{yy}}\rangle = m_0 \,, \quad
\langle J^t \rangle = \mu r_0 \,, \quad \langle O_1 \rangle =0 \,,
\end{equation}
with all others vanishing. {$\langle{T^{tt}}\rangle  = 2 \langle{T^{xx}}\rangle$ implies that charge carriers are still of massless character.}

Now we want to study the responses of this system for small perturbations.
In particular we are interested in the electric conductivity, which is related to the boundary current operators $ \vec{J}$.  Because of rotational symmetry in $x$-$y$ space, it is enough to consider $ J_x$.  Since this operator is dual to the bulk gauge fields $A_x$,  we consider a following linear fluctuation around the background
\begin{align} \label{flucA}
 \delta A_x(t,r) &= \int^{\infty}_{-\infty} \frac{\dd \omega}{2\pi}  e^{-i\omega t}  a_{x}(\omega,r) \,.
\end{align}
The fluctuation is chosen to be independent of $x$ and $y$.
It is allowed since all the background fields entering the equations of motion are independent of $x$ and $y$.
The gauge field fluctuation($\delta A_x(t,r)$) turns out to source metric($\delta  g_{tx}(t,r) $) and scalar field($\delta \psi_1(t,r)$) fluctuation
\begin{align}
\delta g_{tx}(t,r) &=  \int^{\infty}_{-\infty} \frac{\dd \omega}{2\pi} e^{-i\omega t} \frac{r^2}{r_0^2} h_{tx}(\omega,r),  \label{flucG} \\ \label{flucPsi}
 \delta \psi_1(t,r) &= \int^{\infty}_{-\infty} \frac{\dd \omega}{2\pi} e^{-i\omega t}  \chi (\omega,r) \,,
\end{align}
and all the other fluctuations can be decoupled.
Since we will work in momentum space, we defined the momentum space functions $a_x, h_{tx}$, and  $\chi$\footnote{$\chi$ here is the same as $\alpha^{-1} \chi$ in \cite{Andrade:2013gsa}.}, where $h_{tx}(\omega, r)$ is defined so that it goes to constant as $r$ goes to infinity.

In momentum space, the linearised  equations  around the background  are %
\begin{align}
\frac{\beta ^2  h_{tx}}{r^2 f}+\frac{i r_0^2 \beta  \omega   \chi}{r^2 f}-\frac{\mu r_0^3  a_x'}{r^4}-\frac{4  h_{tx}'}{r}-h_{tx}''=&0 \,, \label{eq1}\\
\frac{i \beta r_0^2 f  \chi'}{r^2 \omega }+\frac{\mu r_0^3  a_x}{r^4}+ h_{tx}'=&0\,, \label{eq2}\\
\frac{f'  a_x'}{f}+\frac{\mu   h_{tx}'}{ r_0 f}+\frac{\omega ^2  a_x}{f^2}+ a_x''=&0\,, \\
\frac{f'  \chi'}{f}-\frac{i \beta  \omega   h_{tx}}{r_0^2 f^2}+\frac{\omega ^2  \chi }{f^2}+\frac{2  \chi'}{r}+ \chi''=&0\,. \label{eq4}
\end{align}
which are obtained from \eqref{Rmn}-\eqref{Epsi}.
Among these four equations, only three are independent.\footnote{{The equations \eqref{eq1}-\eqref{eq4} may be decoupled in terms of gauge invariant combinations~\cite{Andrade:2013gsa}. The equation governing electric conductivity turns out to be the same as the one in the massive gravity model \cite{Davison:2013jba}, while the equations for  thermal/thermoelectric conductivities are different.}}   We need to solve these equations satisfying two boundary conditions: incoming boundary conditions at the black hole horizon and the Dirichlet boundary conditions at the boundary.
Near the boundary ($r \rightarrow \infty$) the asymptotic solutions read
\begin{equation} \label{nearb}
a_x=a_x^{(0)} + \frac{1}{r}a_x^{(1)}+ \cdots,
\end{equation}
and the DC electric conductivity \cite{Andrade:2013gsa}  is
\begin{equation} \label{DC1}
\sigma_{DC} 
 =  1+\frac{\mu^2}{\beta^2}  \,,
\end{equation}
which was computed at the horizon (not at the boundary) by rewriting the DC conductivity in terms of a $r$-independent combinations of $a_x$ and $\chi$.  We refer to \cite{Andrade:2013gsa} for details. This technique using $r$-independent quantity is in line with \cite{Iqbal:2008by, Blake:2013bqa}, but does not work for finite $\omega$. To compute AC conductivity we rely on a numerical method, which is the subject of the following section.

\section{General numerical methods with constraint}\label{method}

The analytic method used in \cite{Andrade:2013gsa} is efficient to obtain the DC electric conductivity. However, to compute AC electric conductivity together with AC thermal/thermoelectric conductivity we have to resort to a numerical method. Since the conductivities are related to the retarded Green's functions through the Kubo formula, we need to obtain an action(generating functional) including two sources.
A natural holographic starting point is the on-shell renormalised action to quadratic order in fluctuation fields~\cite{Son:2002sd, Son:2006em, Hartnoll:2009sz}.
 In momentum space the on-shell action with the fluctuations \eqref{flucA}-\eqref{flucPsi} reads
\begin{equation}  \label{S2}
S_{\mathrm{ren}}^{(2)}= \lim_{r\rightarrow \infty} \frac{V_2}{2} \int \dd \omega \left[ - m_0\, h_{tx} h_{tx} - \mu\, a_{x} h_{tx}   -  f(r)  a_{x}a_{x}'   + r^4   h_{tx}  h_{tx}'-  r^2f(r)     \chi  \chi' \right] \,,
\end{equation}
 which is derived from
\begin{align}
\frac{V_2}{2}   \int dt   & \{ -\left(2 r^3+\frac{r^4 f'(r)}{f(r)}+\frac{\beta ^2 r^2-4 r^4}{\sqrt{f(r)}}\right)h_{{tx}}^2-\frac{\beta  r^2  }{\sqrt{f(r)}}\chi  \dot{h}_{{tx}}    \nonumber
\\ &  -h_{{tx}} \left(\mu  \mathit{a}_x-\frac{\beta  r^2 \dot{\chi }}{\sqrt{f(r)}}\right)-r^2 f(r)  \chi \chi '+\frac{r^2  }{\sqrt{f(r)}}\chi  \ddot{\chi }-f(r) \mathit{a}_x \mathit{a}_x'+r^4 h_{{tx}} h_{{tx}}'   \}
\end{align}
where $V_2$ is the two dimensional spatial volume $\int \dd x \dd y$.

Notice that the boundary term at the horizon is deleted according to the prescription to the retarded green function \cite{Son:2002sd}. 
The boundary values of the fields are interpreted as the sources of some dual field theory operators, so we may readily read off the two point functions from the first two terms in \eqref{S2}, while the other three terms look not straightforward.  However, thanks to linearity of equations \eqref{eq1}-\eqref{eq4}, we can always find out the linear recurrence relation between the modes of fields  ($a_{x},~h_{tx},~\chi$). By this linear relation the action is reduced to the schematic form as follows.
\begin{equation}
S_{\mathrm{ren}}^{(2)} = \frac{V_2}{2}  \int \frac{\dd  \omega}{(2\pi)}   J_{-\omega}^a  G_{ab}^R J_\omega^b  ,
\end{equation}
where $J_\omega^a$'s denote the boundary values of $a_{x},~ h_{tx},~\chi$ for a given $\omega$.
Thus the remaining technical task is to find out the relation between necessary modes of  ($a_{x}', ~h_{tx}',~\chi'$) and  those of ($a_{x},~ h_{tx},~\chi$).  See \cite{Amado:2009ts,Kaminski:2009dh} for details on the numerical method. Here we modify it a little bit to be more succinct and economical~\footnote{In some cases the equations may be separable in terms of master fields. However, our method applies to any number of coupled fields straightforwardly and we don't need to try to figure out master fields.}.

To develop a systematic method in a general setup let us start with $N$ fields  $\Phi^a(x,r)$, $a=1,2,\cdots, N$,
\begin{equation}
\Phi^a(x,r) = \int \frac{\dd^d k}{(2\pi)^d}  e^{-ikx}  r^q \Phi^a_k(r)\,, \label{newphi}
\end{equation}
where the index $a$ includes components of higher spin fields. $r^q$ is multiplied such that {\it the classical solution of $\Phi^a_k(r) $ goes to constant at boundary}. For example, $q=2$ in \eqref{flucG}.
 A general on-shell quadratic action in momentum space has the form of
\begin{equation} \label{sb}
S_{\mathrm{ren}}^{(2)}  = \lim_{r \rightarrow \infty} \frac{1}{2} \int \frac{\dd^d k}{(2\pi)^d}  \left[ \Phi_{-k}^a(r) \mathbb{A}_{a b}(r,k) \Phi_k^b(r)
+  \Phi_{-k}^a(r) \mathbb{B}_{a b}(r,k) \partial_r{\Phi_k^b}(r)  \right],
\end{equation}
where $\Phi^a_k(r) $ are solutions of  linear second order differential equations of $N$ fields $\Phi^a_k(r) $.  $\mathbb{A}$ and $\mathbb{B}$ are regular matrices of order $N$. The renormalized action (\ref{sb}) is assumed to contain all the counter terms. For example, see \eqref{S2} for an action and \eqref{eq1}-\eqref{eq4} for a system of equations.

When  the differential equations are second order we need to give $2N$ boundary conditions: $N$ of them are at the horizon (call them $\phi^a$, $a=1, ..., N$)  and other  $N$ (call them $J^a$)  are at the boundary.  For numerical integration, we have to convert the boundary value problem to 
the initial value problem, by considering $2N$ canonical initial data at the event horizon. We solve the initial value problem for $N$  independent initial value set
and judiciously combine $N$ set of  solutions such that final value of solution  is identical to the boundary value we have chosen. The procedure will be independent of the chose of canonical initial data as we will show below. 
 
Near horizon($r=1$), solutions can be expanded as
\begin{equation} \label{incoming}
\Phi^a(r) = (r-1)^{\nu_{a\pm}} \left( \varphi^{a} + \tilde{\varphi}^{a} (r-1) + \cdots \right)
\end{equation}
where we omitted the subscript $k$ for simplicity and $\pm$ correspond to incoming/outgoing boundary conditions. To compute the retarded Green's function we choose the incoming boundary condition \cite{Son:2002sd},  fixing $N$ initial  conditions. The other $N$ initial conditions correspond to $N$ independent sets of $\varphi^{a}$, denoted by $\varphi^{a}_{i}$, $i=1,2,\cdots, N$.  Notice that due to incoming boundary condition, $\varphi^a$ determines $\tilde{\varphi}^a$ through horizon-regularity condition so that we can determine the solution completely. 

For example, $\varphi^{a}_{i}$ may be chosen as
\begin{equation} \label{init}
\begin{pmatrix}
    \varphi^{a}_{1} \ & \varphi^{a}_{2}\ & \varphi^{a}_{3}\ &  \ldots \ &  \varphi^{a}_{N}
\end{pmatrix}
   =
\begin{pmatrix}
    1 & 1& 1&  \ldots & 1 \\
    1 & -1& 1 & \ldots & 1 \\
   1 & 1& -1 & \ldots & 1 \\
    \vdots & \vdots & \vdots  & \ddots & \vdots \\
    1 & 1 & 1 & \ldots & -1
\end{pmatrix}
\end{equation}
where  $\phi^a_j$ {with  fixed } j $ (\ge 2)$ is a column vector  $\vec{\phi}_j$ whose
$a$-th  element is $-1$ if  $a=j$ and $1$ otherwise. 
  For the case without constraint equation,  each  initial value column $\vec{\phi}_j$   yields a   solution, 
  which will be denoted by $\vec{\Phi}_j(r)$. 
  Since all fields have been redefined such that $\Phi^a$ goes to constant values at the boundary in \eqref{newphi}, 
near boundary ($r \rightarrow \infty$) the solution is expanded as\footnote{If we simply expand the solution near boundary there may be terms of lower power than $1/r^{\delta_a}$. For example, see $ g^{(2)}_{\mu\nu}$ and $\psi^{(2)}_I $ in \eqref{nearb0}. Those terms can be taken care of by counter terms and we don't write them here to focus on essential ideas.}
\begin{equation}
\Phi_i^a(r)  \rightarrow   \mathbb{S}_{i}^{a}  + \frac{\mathbb{O}_{i}^{a}}{r^{\delta_a}}  + \cdots   \qquad (\mathrm{near\ boundary})\,,
\end{equation}
where  $\mathbb{S}_i^a$ denote the   leading terms of $i$-th  solution  and  $\mathbb{O}_{i}^{a}$ are the  sub-leading term  determined by 
   horizon data, namely by initial conditions $\phi_i^a$.  Notice that $\mathbb{S}$ and $\mathbb{O}$ can be written as regular matrices of order $N$, where the superscript $a$ runs for row index and the subscript $i$ runs for column index.

The general solution is a linear combination of them: let 
\begin{equation} \label{GS} 
\Phi^a(r) = \Phi_{i}^{a}(r) c^i
\end{equation}
with  real constants $c^i$'s. We need to choose $c^i$ such that the combined source term matches the boundary value $J^a$: 
\begin{equation} \label{source}
J^a = \mathbb{S}_{i}^{a} c^i \,.
\end{equation} 
Our aim is to read off the response of $\partial_r{\Phi^a}(r) \sim \mathbb{O}^a_i c^i$ with respect to $J^a$. When there is no constraint related to the diffeomorphic invariance, it can be done simply by noting that $c^i$ is expressed in terms of $J^a$
\begin{equation}
 c^i  = (\mathbb{S}^{-1})^{i}_{a} J^a \,.
\end{equation}
In this case, the second term of \eqref{sb} may be written as
\begin{equation} \label{response}
\mathbb{B}_{ac}(r,k)\partial_r{\Phi^c}(r)    =  \sum_{c,i} \left[-\mathbb{B}_{ac} (r,k) (\delta_c r^{-\delta_c -1}\mathbb{O}_{k}^{c} ) (\mathbb{S}^{-1})^{i}_{b} \right] J^b + \cdots := \left[\mathbb{C}_{ab}(r,k)\right] J^b +\cdots\,.
\end{equation}
Notice that $\mathbb{C}_{ab}(\infty,k)$ is a finite value  because $\mathbb{B}_{ac} (r,k)  \sim r^{\delta_c +1}$ as $r \rightarrow \infty$. 
The essential structure of the matrix $\mathbb{C}_{ab}$ is the multiplication of three matrices $\mathbb{B}\cdot \mathbb{O} \cdot \mathbb{S}^{-1}$, which  shows the independence of the choice of the initial condition \eqref{init}, because the different choice of initial value vectors are nothing but a linear transformation $ {\phi^a_i} \to {\phi^a_j}R^j_i$, which induces right multiplications in the solutions: 
$\mathbb{S} \to \mathbb{S} R, \mathbb{O} \to \mathbb{O} R $.  

With \eqref{source} and \eqref{response} the final boundary action yields
\begin{equation}
S_{\mathrm{ren}}^{(2)}  = \frac{1}{2}  \int \frac{\dd^d k}{(2\pi)^d}  \left[ J_{-k}^a \left[\mathbb{A}_{a b}(\infty,k) + \mathbb{C}_{ab}(\infty,k)\right]J_k^b  \right],
\end{equation}
where we reinserted the subscript $k$. Since matrices $ \mathbb{A}, \mathbb{C}$ are independent of $J$, the retarded Green's function is
\begin{equation} \label{Green0}
G_{ab}^R = \mathbb{A}_{a b}(\infty,k) + \mathbb{C}_{ab}(\infty,k) \,.
\end{equation}
Notice that 
for one field case without mass term, this is the well known structure of the retarded Green's function: $\mathbb{A}=0$ and $ G^R \sim \mathbb{O}/\mathbb{S}$, that is, the green function is  the ratio of the coefficients of  the subleading term and the leading term.

In summary, to compute the retarded Green's function we need four square matrices of order $N$(the number of fields): $\mathbb{A}, \mathbb{B}, \mathbb{S}, \mathbb{O}$. $\mathbb{A}$ and $\mathbb{B}$ can be read off from the boundary action \eqref{sb}.
$\mathbb{S}$ and $\mathbb{O}$ are given  from the solution of a set of differential equations. We have to solve $N$ times with $2N$ independent initial conditions  to construct regular matrices of order $N$.  The retarded Green's function is schematically $\mathbb{A} + \mathbb{B}\cdot \mathbb{O} \cdot \mathbb{S}^{-1} \equiv \mathbb{A} + \mathbb{C}$. The precise form of $\mathbb{C}$ is shown in \eqref{response}.

Our story so far is for the system without constraint. In actual calculation, Einstein equation always contains constraint equations (CE) due to the residual diffeomophism invariance of the linearised equation of motion. 
We describe  how to fix this complication for the case 
of one CE for notational convenience. Generalization to two or more CE is straightforward.   
At the horizon, the CE relates the (initial) field values. 
For example, the last component of $\vec{\varphi}_j$  can be determined  by other components. So the space of initial value vectors (IVV)     is N-1 dimensional subspace. 
Solving differential equation using such N-1  IVV gives, of course, only N-1 solutions.  
However, when we give boundary conditions, 
 we formally assign N  boundary values  $J_{i}$.  
As a consequence, eq. (\ref{source}) is not invertible!
To fix this problem, we introduce a vector along the gauge orbit direction,
$\vec{S}_0= \delta_\xi \Phi(\infty)$   to the space spanned by  $\{ \vec{S}_1, \cdots, \vec{S}_{N-1}  \}$. 
Then, 
\begin{equation} \label{reconst}
\sum_{j=1}^{N-1} \vec{S}_j c^j + \vec{S}_0 ~ c^0 =\vec{J}.
\end{equation} 
Now this equation is invertible. We can find proper $c^0$ and $c^i$'s for any choice of initial condition vectors to fit the given $\vec{J}$. 
Notice that $\vec{S}_0$ does not generate true degree of freedom  since 
it is along the diffeomorphism orbit direction.  
It is not hard to see  why $\vec{S}_0$ actually satisfies the on-shell condition:
This happens since the residual gauge  transformation  leaves the linearised equation of motion invariant:  if we write the equation of motion as
$\mathbb{M}\cdot \Phi(r)=0$, then for a diffeomorphism that preserve the 
equation of motion, $\delta \mathbb{M}=0$.  Then by taking the gauge variation of the equation of motion,  we have 
$\mathbb{M}\cdot \delta \Phi=0$.
 Although  gauge variation is not a true solution satisfying in falling conditions, 
 it can serve as an element of a basis of the on-shell $J$ space and on shell condition.  
  
Furthermore, 
change of initial vectors $\vec{\varphi_j}$ by $\phi^{a}_{i}\to  \varphi^{a}_{j}R^{j}_{i}$  which results  in  $S^{a}_{j}R^{j}_{i}$ can be undone by changing the 
$c^{i}\to c^{j}(R^{-1})_{j}^{i}$, where all the indices run from 1 to $N-1$.  
Notice that in this proof of basis independence, we used the fact that   the differential equations define  a linear map $U(r):  \varphi^a_i \to U^a_b (r)  \varphi_i^b$, so that the evolution operator $U(r)$  is a left multiplication (acting on upper index $a$), while basis change is a right multiplication acting on index $i$, so that  initial  data  ${\varphi}^{a}_{i}$ at the horizon  and the final  solution  ${S}^{a}_{i}$ at the boundary is multiplied by the same matrix $R$.   

 
\begin{figure}[]
\centering
  \subfigure[Re $\sigma$.  Delta functions at $\omega=0$ are not drawn. The red dots at $\omega = 0$ are the analytic values $\sigma_Q$ in \eqref{Jo}.]
   {\includegraphics[width=4.6cm]{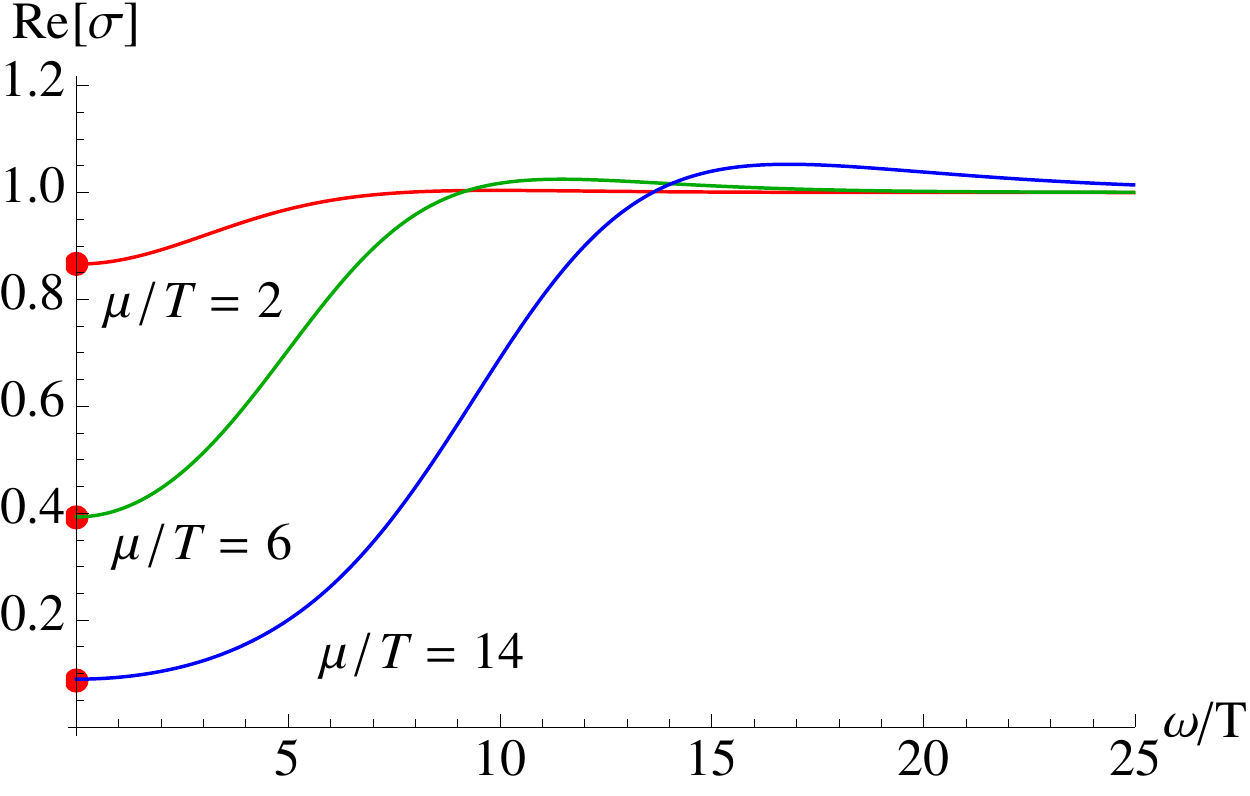} \label{}}  \hspace{2mm}
     \subfigure[Im $\sigma$. There are $1/\omega$ poles corresponding to delta functions in (a)]
   {\includegraphics[width=4.6cm]{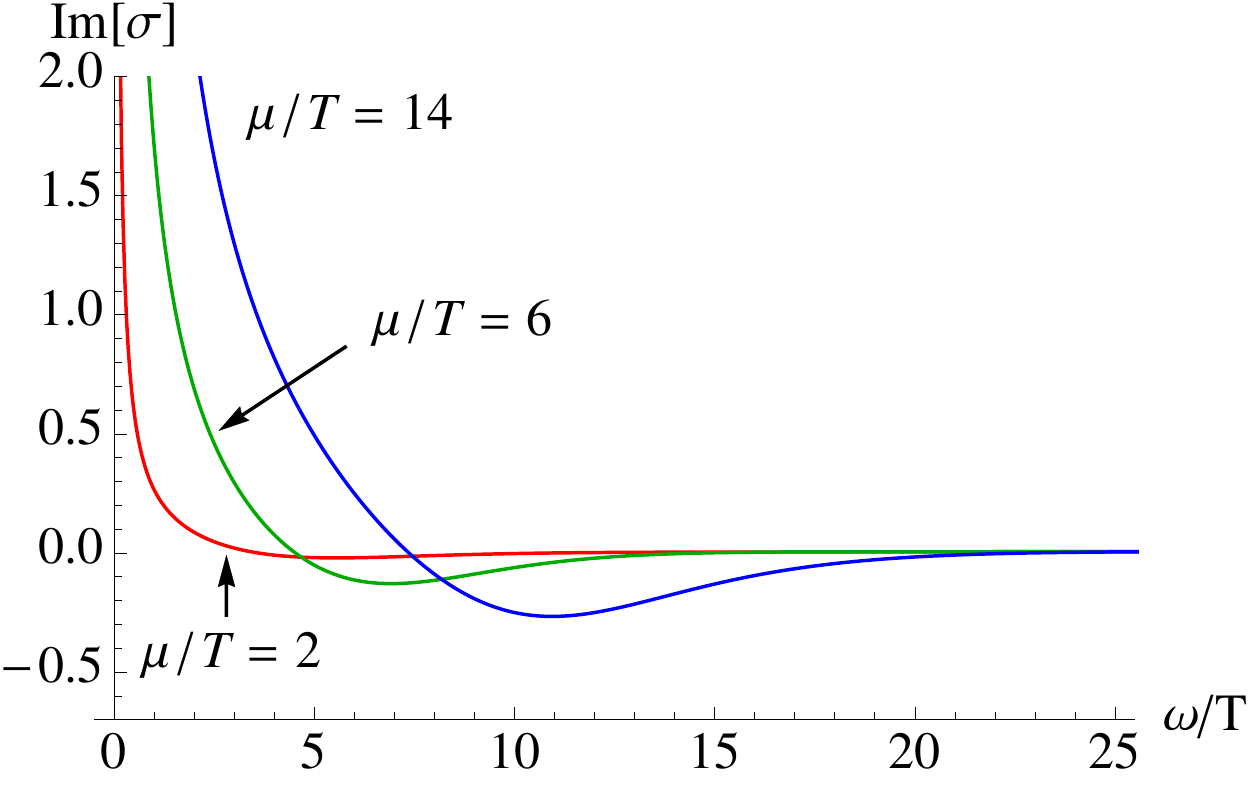} \label{} }  \hspace{2mm}
     \subfigure[$\omega$ Im $ \sigma$. The red dots at $\omega = 0$ are the analytic values $K$ in \eqref{Jo}.]
     {\includegraphics[width=4.6cm]{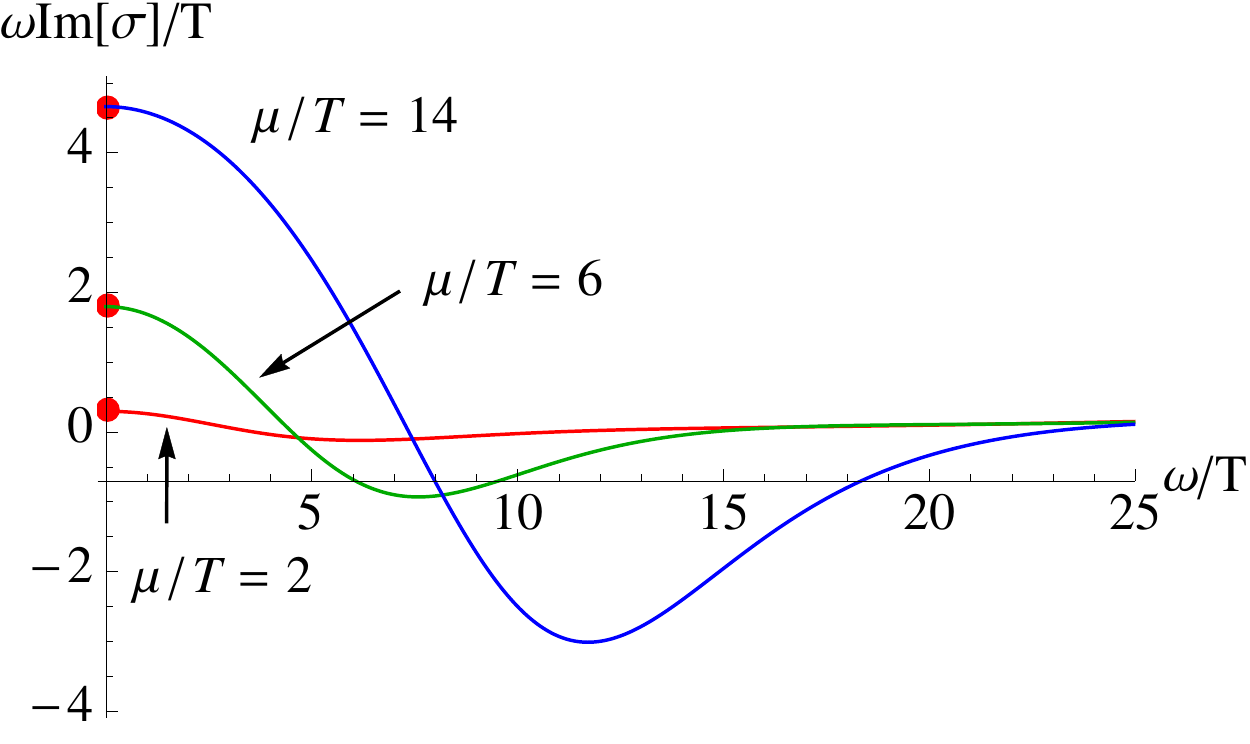} \label{}}
 \caption{Electric conductivity without momentum relaxation ($\beta=0$). } \label{beta0}
\end{figure}

In order to check the validity of our numerical method and code, we computed AC electric conductivity when $\beta=0$. Our numerical plot is shown in Figure \ref{beta0}, which agrees to the Figure 6 of \cite{Hartnoll:2009sz}.  It is a nontrivial consistency check of our method since the plot in  \cite{Hartnoll:2009sz} has been obtained by solving a single equation of the gauge field $a_x$, while we have solved coupled equations of $a_x$ and $g_{tx}$.  Of course if the coupled equations can be decoupled as shown in \cite{Hartnoll:2009sz} there is no point of solving coupled equations. However, because this decoupling is not always possible it is important to develop a systematic and efficient method for coupled fields cases.  In addition to the agreement of Figure \ref{beta0} to Figure 6 of \cite{Hartnoll:2009sz} our results in $\omega \rightarrow 0$ limit match the analytic form in \cite{Hartnoll:2007ip,Ge:2010yc,Matsuo:2009yu,Ge:2008ak}
\begin{equation} \label{Jo0}
\sigma = \sigma_Q + i\frac{K}{\omega} \,,
\end{equation}
where
\begin{equation} \label{Jo}
\sigma_Q = \left(\frac{3-\frac{\mu^2}{4 {r}_0^2}}{3+\frac{3\mu^2}{4{r}_0^2}}\right)^2 \,,\qquad
K = {r}_0\frac{\frac{\mu^2}{{r}_0^2}}{3+\frac{3\mu^2}{4{r}_0^2}} \,, 
\end{equation}
and ${r}_0$ is defined by \eqref{r0} evaluated at $\beta=0$.  
The red dots at $\omega=0$ in Figure \ref{beta0}(a) and (c) are analytic values $\sigma_Q$ and $K$  respectively.
To read off $K$ we made a plot for $\omega \mathrm{Im} \sigma$ in Figure \ref{beta0}(c) from Figure \ref{beta0}(b). 
  
\section{Electric/thermal/thermoelectric AC conductivities} \label{RN}

In the previous section, we have computed AC conductivity when $\beta=0$ as an exercise.
Now we want to attack our main problem, AC conductivity with moment dissipation generated by $\beta \ne 0$.
The basic setup and relevant equations were introduced in subsection \ref{22}.  We can read off
the conductivities from the action \eqref{S2}.
To closely follow the general methods presented in section \ref{method} we rewrite the action as 
\begin{equation} \label{}
s_{\mathrm{ren}}^{(2)} \equiv \frac{S_{\mathrm{ren}}^{(2)}}{V_2} = \lim_{r \rightarrow \infty} \int \frac{\dd \omega}{2\pi}  \left[ \Phi_{-\omega}^a(r) \mathbb{A}_{a b}(r,\omega) \Phi_\omega^b(r)
+  \Phi_{-\omega}^a(r) \mathbb{B}_{a b}(r,\omega) \partial_r{\Phi_\omega^b}(r)  \right],
\end{equation}
with
\begin{align}
\Phi^a =
\begin{pmatrix}
    a_x  \\
    h_{tx} \\
   \chi \\
\end{pmatrix}\,, \quad
\mathbb{A} = \begin{pmatrix}
    0 & -\mu/2& 0  \\
    -\mu/2 & -m_0& 0  \\
   0 & & 0  \\
\end{pmatrix}\,, \quad
   \mathbb{B}= \begin{pmatrix}
 -f(r) & 0 & 0 \\
 0 & r^4 & 0 \\
 0 & 0 & -r^2 f(r) \\
\end{pmatrix}
\end{align}
where the index $\omega$ is suppressed. In matrix notation, $\Phi^a_{-\omega}$ should be understood as a row matrix.

To compute the matrix $\mathbb{C}$ in \eqref{response} we have to solve the equations \eqref{eq1}-\eqref{eq4}, which we rewrite here setting $r_0=1$:
\begin{align}
\frac{\beta ^2  h_{tx}}{r^2 f}+\frac{i \beta  \omega   \chi}{r^2 f}-\frac{\mu   a_x'}{r^4}-\frac{4  h_{tx}'}{r}-h_{tx}''=&0 \,, \\
\frac{i \beta  f  \chi'}{r^2 \omega }+\frac{\mu   a_x}{r^4}+ h_{tx}'=&0\,, \\
\frac{f'  a_x'}{f}+\frac{\mu   h_{tx}'}{f}+\frac{\omega ^2  a_x}{f^2}+ a_x''=&0\,, \\
\frac{f'  \chi'}{f}-\frac{i \beta  \omega   h_{tx}}{f^2}+\frac{\omega ^2  \chi}{f^2}+\frac{2  \chi'}{r}+ \chi ''=&0\,,
\end{align}
Since only three equations are independent we may solve any three of them.
Near the black hole horizon ($r \rightarrow 1$) the solutions are expanded as
\begin{equation} \label{nearh}
\begin{split}
&h_{tx} =  (r-1)^{\nu_\pm+1}(h_{tx}^{(I)} + h_{tx}^{(II)}(r-1) + \cdots),  \\
&a_x=  (r-1)^{\nu_\pm}(a_x^{(I)} + a_x^{(II)}(r-1) + \cdots )   ,\\
&\chi=  (r-1)^{\nu_\pm}(\chi^{(I)} + \chi^{(II)}(r-1) + \cdots )
\end{split}
\end{equation}
where $\nu_\pm = \pm i 12 \omega  /(-12  + 2\beta^2 + \mu^2)$ and the incoming boundary condition corresponds to $\nu = \nu_+$.   
Near the boundary ($r \rightarrow \infty$) the asymptotic solutions read
\begin{equation} \label{nearb}
\begin{split}
h_{tx} &=   h^{(0)}_{tx} + \frac{1}{r^2} h^{(2)}_{tx} + \frac{1}{r^3}h_{tx}^{(3)}+\dots,  \\
a_x&=a_x^{(0)} + \frac{1}{r}a_x^{(1)}+ \cdots, \qquad  \quad \\
 \chi &= \chi^{(0)} + \frac{1}{r^2} \chi^{(2)}+ \frac{1}{r^3}\chi^{(3)} + \cdots 
\end{split}
\end{equation}
With incoming boundary condition and initial values \eqref{init} at horizon  we numerically integrate the equations from the horizon. 

 For our equations there is one subtlety caused by a symmetry of the system.
Analysing the equations near the {\it horizon} with the expansion \eqref{nearh} we find that only two of $a_x^{(I)} $, $\chi^{(I)}$, and $h_{tx}^{(I)}$ are free,
which is due to the gauge fixing $g_{rx}=0$. 
Therefore, we don't have a complete basis   to reconstruct a general 
boundary value vector $\vec{J}$.  
However, there is a residual gauge transformation keeping $g_{rx}=0$, which is generated by the vector field 
$\xi^\mu$ whose non-vanishing component is $\xi^x= \epsilon e^{-i\omega t} $. So we add a vector along the residual gauge orbit. 
Since  ${\cal L}_\xi g_{tx}=-i\omega r^2 \xi^x$ and  
${\cal L}_\xi \varphi = \beta \xi^x$,  
${\cal L}_\xi \Phi  = (0, -i\omega \xi^x, \beta \xi^x)^T$, independent of $\Phi$.  Therefore  we can choose 
 \begin{equation}\label{constant1}
\vec{S}_0 = (a_x^{0},  h_{tx}^0, \chi^0)^T =( 0,  1, i\beta/\omega)^T . 
\end{equation}  
Notice  that  $ \vec{S}_0$   satisfies the equations of motion, 
 as we mentioned in earlier section. Since $ \vec{S}_0$ happened to be r-independent, it is equivalent to formally add a  `constant'  solutions
\begin{equation} \label{constant2}
a_x = 0 \,, \quad  h_{tx} = h_{tx}^0 \,, \quad \chi = \frac{i \beta h_{tx}^0}{\omega} \,,
\end{equation}
where $h_{tx}^0$ are arbitrary constants.  
This kind of solution for the theory with $\beta=0$ was introduced in 
\cite{Hartnoll:2007ai}. 
It is interesting that \eqref{constant2} is similar to (3.8) of 
\cite{Donos:2013eha}, which is  a condition imposed at infinity to extract the gauge-invariant conductivity.\footnote{We thank Aristomenis Donos and Jerome Gauntlett for pointing out this similarity.}
They first choose initial data such that the final solution lies on the gauge orbit and 
then using the gauge transformation, one can set $\chi^{0}=  h_{tx}^0=0$. 
The  eq. $\chi^{(0)}- \frac{i\beta}{\omega}h^{(0)}_{tx}=0$ 
   is nothing but the equation of  the gauge orbit (in the space of  leading component of solutions) passing $(a_x^{0} , 0,0)$.

Now,  $a_x^{(1)},\chi^{(3)}, h^{(3)}_{tx}$ 
are in general  linear combinations of $a_x^{(0)},\chi^{(0)}, h^{(0)}_{tx}$ due to the boundary conditions and in the above choice of solution and  gauge where $  \chi^{(0)}=0=h^{(0)}_{tx}$, 
  one can calculate electric conductivity by $a_x^{(1)}/a_x^{(0)}$.  In this approach, one can calculate only electric conductivity and shooting is cumbersome. 
  
In our method  generic solution $\vec{S}_1, \vec{S}_2$   satisfy $S_{i3}\ne \frac{i\beta}{\omega}S_{i2} $ for $i=1,2$,  hence  $\vec{S}_1, \vec{S}_2, \vec{S}_0$ form a  basis to make 
$ \vec{S}_1 c^1 + \vec{S}_2 c^2+ \vec{S}_0  c^0 =\vec{J}$ invertible.
All the green functions are calculated simultaneously.  
Again, we emphasise that true physical degree of freedom is 2 dimension 
and $\vec{S}_0$ does not generate a true physical degree of freedom.

\subsection{Green functions and Transport coefficients}
Having computed numerically the matrices $\mathbb{S}$ and $\mathbb{O}$, we may construct a $3 \times 3$ matrix of retarded Green's function. We will focus on the $2 \times 2$ submatrix corresponding to $a_x^{(0)}$ and $h_{tx}^{(0)}$ in \eqref{nearb}. Since $a_x^{(0)}$ is dual to $U(1)$ current $J_x$ and $h_{tx}^{(0)}$ is dual to energy-momentum tensor $T_{tx}$ the Green's function matrix may be written as
\begin{equation}
\label{Green1}
\left(\begin{array}{cc}   G^R_{J_xJ_x} & G^R_{J_xT_{tx}} \\   G^R_{T_{tx}J_x}  &  G^R_{T_{tx}T_{tx}} \end{array}\right) =: \left(\begin{array}{cc}   G_{11} & G_{12} \\   G_{21}  &  G_{22} \end{array}\right)~,
\end{equation}
where we introduced the second term for notational simplicity.
From the linear response theory,  we have the following relation between the response functions and the sources:
\begin{equation}
\label{theo}
\left(\begin{array}{c}\langle J_x \rangle \\ \langle T_{tx} \rangle \end{array}\right)=
\left(\begin{array}{cc}   G_{11} & G_{12} \\   G_{21}  &  G_{22} \end{array}\right)
\left(\begin{array}{c}  a_x^{(0)} \\  h_{tx}^{(0)}\end{array}\right)\,.
\end{equation}

We want to relate the Green's functions \eqref{Green1} to phenomenological transport coefficients.
Our goal is to study the electric, thermal, thermoelectric conductivities defined as
\begin{equation}
\label{pheno}
\left(\begin{array}{c}\langle J_x \rangle \\ \langle Q_x \rangle \end{array}\right)
=
\left(\begin{array}{cc}   \sigma & \alpha T \\   \bar{\alpha} T & \bar{\kappa} T \end{array}\right)
\left(\begin{array}{c} E_x \\ - (\nabla_x T)/T\end{array}\right)~,
\end{equation}
where $\sigma$ is the electric conductivity, $\alpha, \bar{\alpha}$ are the thermoelectric conductivities, and $\bar{\kappa}$ is the thermal conductivity. $Q_x$ is the heat current, $E_x$ is an electric field and $\nabla_x T$ is a temperature gradient. By taking into account a diffeomorphism invariance \cite{Hartnoll:2009sz,Herzog:2009xv}, \eqref{pheno} can be expressed as
\begin{equation}
\label{pheno1}
\left(\begin{array}{c}\langle J_x \rangle \\ \langle T_{tx} \rangle -\mu \langle J_x \rangle \end{array}\right)
=
\left(\begin{array}{cc}   \sigma & \alpha T \\   \bar{\alpha} T & \bar{\kappa} T \end{array}\right)
\left(\begin{array}{c} i\omega  ( a_x^{(0)} + \mu  h_{tx}^{(0)})   \\   i \omega  h_{tx}^{(0)} \end{array}\right).
\end{equation}
Comparing  \eqref{pheno1} and \eqref{theo} we have
\begin{equation} \label{pheno2}
\left(\begin{array}{cc}   \sigma & \alpha T \\   \bar{\alpha} T & \bar{\kappa} T \end{array}\right) =
\left(
\begin{array}{cc}
 -\frac{i G_{11}}{\omega } & \frac{i (G_{11} \mu -G_{12})}{\omega } \\
 \frac{i (G_{11} \mu -G_{21})}{\omega } & -\frac{i (G_{22} -`G_{22}(\omega=0){\textrm'}+\mu  (-G_{12}-G_{21}+G_{11} \mu ))}{\omega } \\ 
\end{array}
\right) .
\end{equation}
Notice that the term in quotation marks $`G_{22}(\omega=0)\textrm'$ is subtracted 
according to the linear response theory: one has to subtract static term whenever the latter is non-zero \cite{Kovtun:2012rj}.  This is related to the fact that for the  causal Green functions, we have to  apply source fields only in the past not in future. 
Without such subtraction, we would have unphysical pole in $\Im[\bar \kappa]$. \footnote{We thanks the  referee to point this out. }  Some authors prefer to  eliminate such pole by energy dependent counter term \cite{Amoretti:2014zha}.   
 In summary, we numerically compute $G_{11}, G_{12}, G_{21}, G_{22}$ by \eqref{Green0} and combine them as \eqref{pheno2} for physical conductivities.

\subsection{Optical conductivity and coherent/incoherent metal}

\begin{figure}[]
\centering
  \subfigure[Re $\sigma$. A delta function at $\omega=0$ for $\beta=0$ is not drawn. The red dots at $\omega = 0$ are the analytic DC values \eqref{DC2}.  ]
   {\includegraphics[width=6cm]{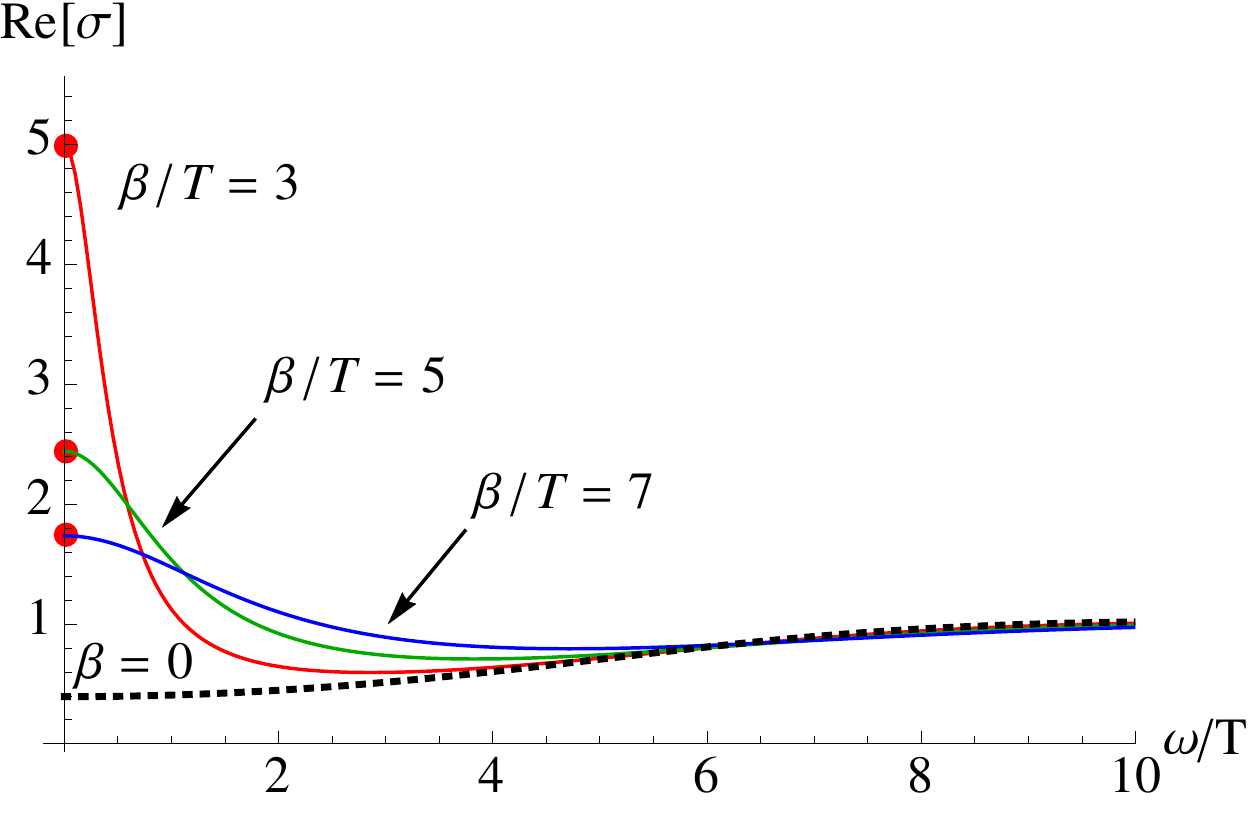} \label{sigmaa}} \hspace{5mm}
     \subfigure[Im $\sigma$. There is a $1/\omega$ pole for $\beta=0$ corresponding to a delta function in (a)]
   {\includegraphics[width=6cm]{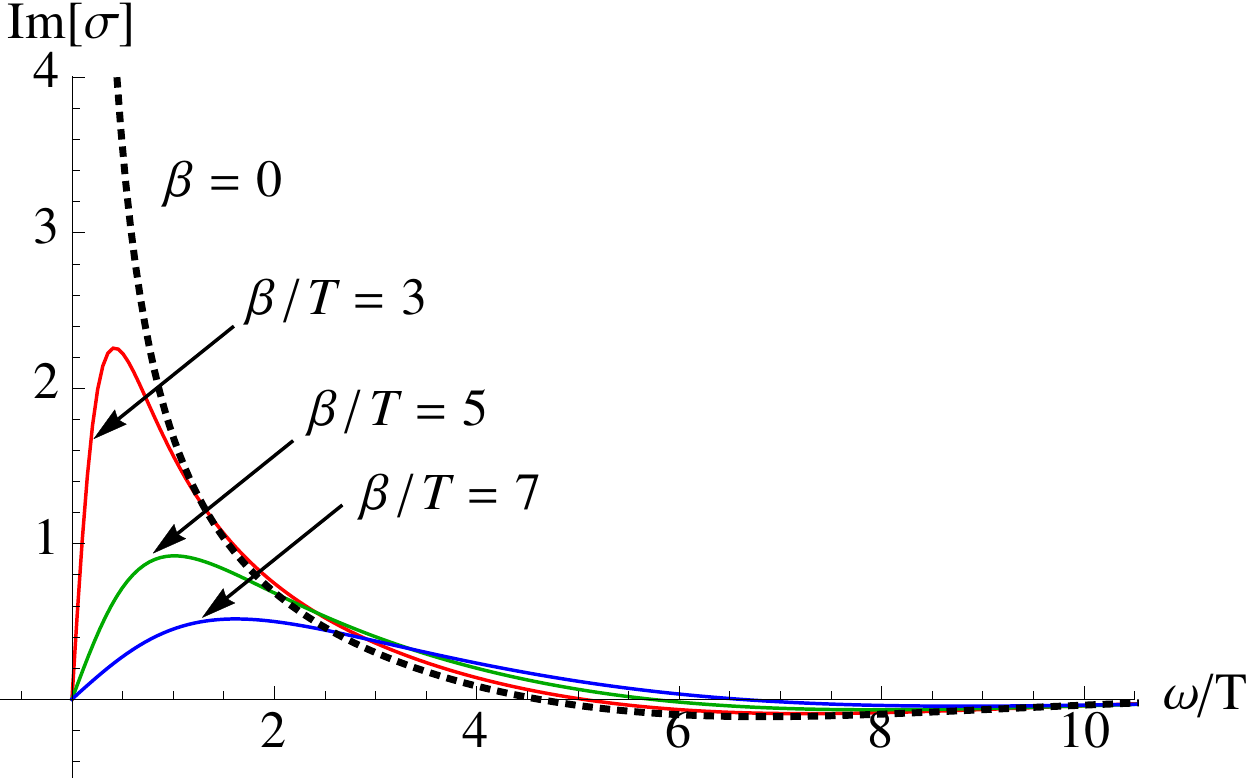}\label{sigmab} }
 \caption{Electric conductivity $\sigma$ with momentum relaxation at fixed $\mu/T = 6 $.
 For larger $\beta$ the Drude-like peak  at small $\omega$ becomes broader. As we increase $\beta$, the Drude peak disappears and the transition to incoherent metal is manifest.
           } \label{sigmabeta}
\end{figure}
\begin{figure}[]
\centering
     \subfigure[Re $\sigma$]
     {\includegraphics[width=6cm]{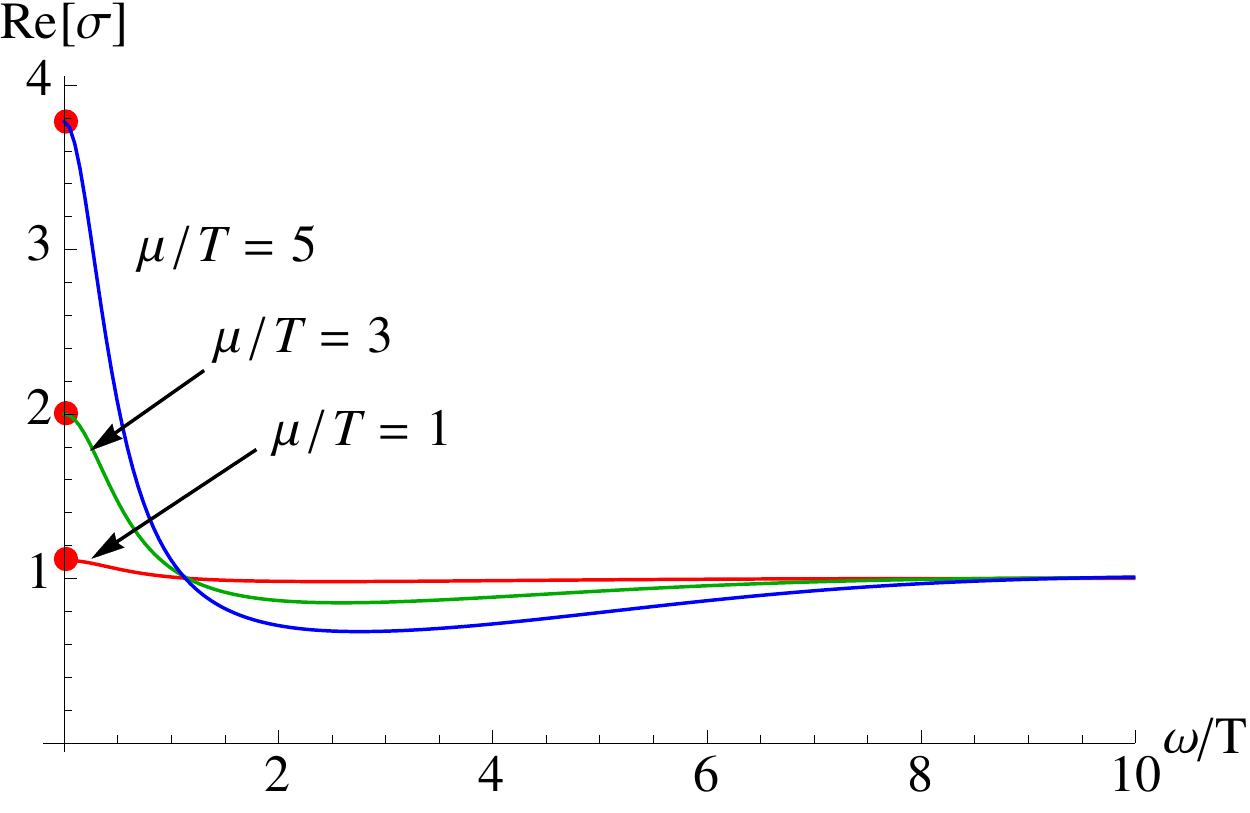} \label{}} \hspace{5mm}
     \subfigure[Im $\sigma$]
   {\includegraphics[width=6cm]{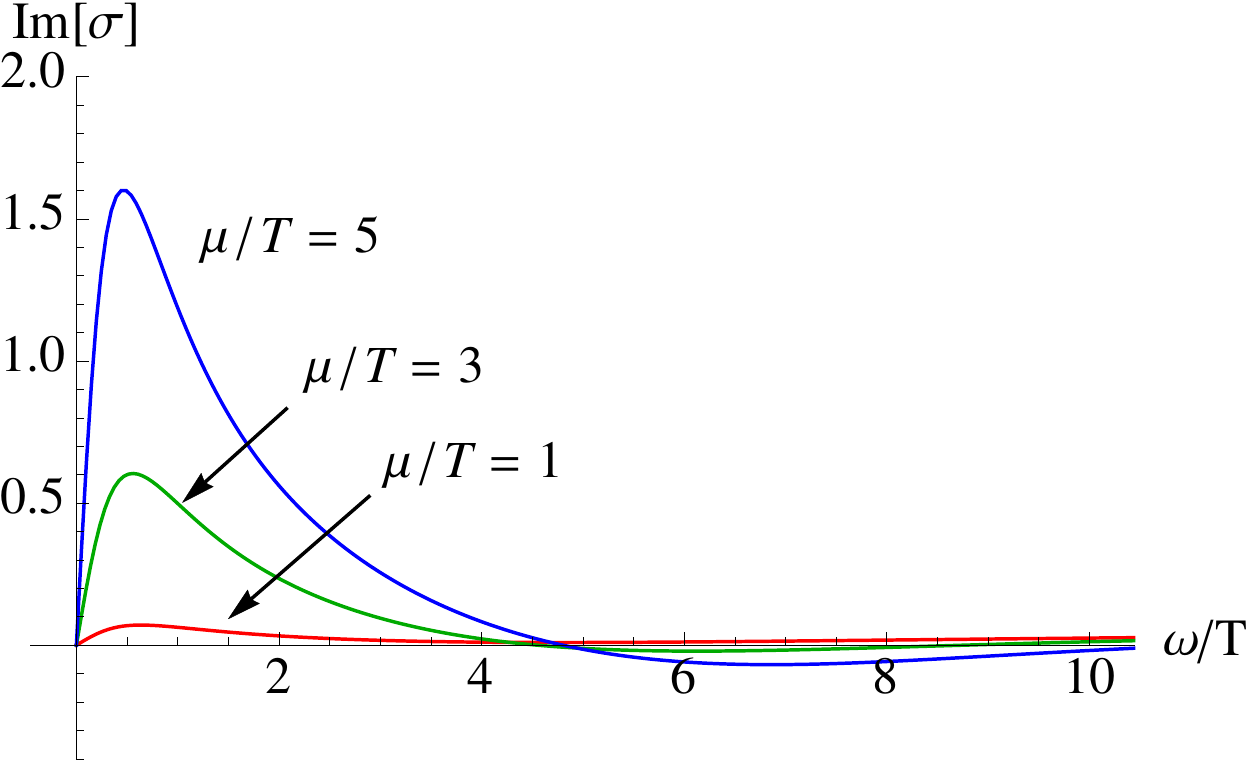} \label{} }
 \caption{Electric conductivity  $\sigma$  with momentum relaxation at fixed $\beta/T = 3$.
 By comparing with Figure \ref{beta0} we may also see how $\beta$ changes conductivity curves since all parameters are the same  except $\beta$. As we decrease $\mu$, the Drude peak disappears and the transition to incoherent metal is manifest.
           } \label{sigmamu}
\end{figure}

In this subsection we present our numerical results on the AC electric conductivity $\sigma$. In Figure \ref{sigmabeta} and \ref{sigmamu}  we focus on the dissipation($\beta$)  effect and the density effect($\mu$) on the AC electric conductivity respectively.  In Figure \ref{taufig} and \ref{Drudes}  we analyse the conductivity at small $\omega$, comparing with the Drude form.  In Figure  \ref{scalings} and \ref{scalingsk} we search for scaling behaviours at intermediate $\omega$.

Figure \ref{sigmabeta} shows how conductivity changes as dissipation strength changes($\beta$).  (a) is the real part and (b) is the imaginary part of the conductivity.  
Two dotted curves($\beta=0$) are the case without momentum dissipation which are the same curves at $\mu/T = 6$ shown in Figure \ref{beta0}.  The coloured solid curves are  the case with momentum dissipation($\beta \ne 0$).
If we turn on a finite $\beta$, a delta function of dotted curve($\beta = 0$) at $\omega=0$ in the real part, which is inferred from $1/\omega$ pole in the imaginary part 
 by Kramers-Kronig relation, becomes a smooth peak with a finite width. 
 At the same time $1/\omega$ pole in the imaginary part disappears.
As $\beta$ increases, the width of the peak in real part increases while the maximum value of the peak (DC conductivities) decreases. In this variation, we checked that the area of the real part of the conductivity does not change numerically. The area difference from $\beta=0$ curve is given by $\frac{K}{2\pi}$, which agrees to the area of the delta function inferred from the imaginary part of the conductivity. This is an example of a sum rule and we have confirmed it for various parameters.  
Numerical DC conductivities agree to the  analytic result \eqref{DC1}
\begin{equation} \label{DC2}
\sigma = 1+\frac{\mu^2}{\beta^2}\,,
\end{equation}
which are shown as the red dots at $\omega = 0$ in Figure \ref{sigmabeta}(a).

Figure \ref{sigmamu} shows the effect of $\mu$ at fixed $\beta$. As $\mu$ increases,  DC conductivity also increases, which is expected since there are more charge carriers. For bigger $\mu$ there is a deeper valley at intermediate $\omega$ regime. This may be expected from Figure \ref{beta0} where a bigger $\mu$ gives a lower value of conductivity at small $\omega$ regime. 

There are two issues on finite frequency regime: one is Drude-like peaks at small frequency and the other is possible scaling laws at intermediate frequency regime. Let us start with an analysis at small frequencies.
The peaks at small $\omega$ in Figure \ref{sigmabeta}(a) and \ref{sigmamu}(a) look similar to Drude peaks qualitatively. For a very small $\beta \ll \mu$, the translation symmetry is broken weakly and we expect to have a Drude form according to \cite{Hartnoll:2012rj}.  For large values of $\beta$ it is possible that the peak is not the standard Drude from.  
As one way to see how much these peaks can resemble  the Drude model, Let us examine the Ward identity. 
At the level of fluctuation the Ward identity \eqref{ward1} is 
\begin{equation}
\partial_t \langle \delta p_{x} \rangle = \beta  \langle \mathcal{\delta O} \rangle + \langle J^t \rangle \delta E_x\,.
\end{equation}
Comparing with the Drude model
\begin{equation}
\frac{dp}{dt} = - \frac{1}{\tau} p + qE \,.
\end{equation}
We see that, if $ \langle \mathcal{\delta O} \rangle$ is proportional to $-\langle \delta p_{x} \rangle$, a Drude-like peak may appear~{\cite{Aprile:2014aja}. 
Furthermore, if $ \langle \mathcal{\delta O} \rangle$ is independent of parameters($\mu, T, \beta$), the scattering time will be inversely proportional to $\beta$. i.e. $\tau \sim 1/\beta$. 
In our case, for $\beta \ll \mu$, it turns out that $ \langle \mathcal{\delta O} \rangle \sim - \frac{\beta}{\mu}\langle \delta p_{x} \rangle  $, 
which will be discussed in \eqref{tauu3}, while, for $ \beta > \mu$, a peak is  different from the Drude form, implying  $ \langle \mathcal{\delta O} \rangle$ is not proportional to $-\langle \delta p_{x} \rangle$.  (see Figure \ref{Drudes} and related discussion).

As a model of peak, let us consider a modified Drude form shifted by $\sigma_Q$
\begin{equation} \label{Drude00}
\sigma(\omega) = \frac{K \tau}{1 - i \omega \tau} + \sigma_Q
\end{equation}
where $\sigma_Q$ is added to take into account the conductivity near 0 frequency at $\beta =0$. 
Since our model is based on AdS-RN black brane solution, there will be a contribution from pair production, $\sigma_Q$, which is also affected by charge density. 
Once we assume \eqref{Drude00}, three parameters $ K, \tau$, and  $\sigma_Q$ can be fixed by considering two limits. \footnote{This form of Drude conductivity implicitly appeared in \cite{Hartnoll:2007ih} with shifted pole due to the magnetic field.}

First, in the limit $\tau \rightarrow \infty$ ($\beta \rightarrow 0$)  $\sigma_Q$ and 
$K$ can be read off from \eqref{Jo}
\begin{equation} \label{Jo0}
\sigma_Q = \left(\frac{3-\frac{\mu^2}{4 {r}_0^2}}{3+\frac{3\mu^2}{4 {r}_0^2}}\right)^2 \,,\quad
K = {r}_0\frac{\frac{\mu^2}{{r}_0^2}}{3+\frac{3\mu^2}{4{r}_0^2}} \,, \,
\end{equation}
where
\begin{equation}
r_0 = \frac{2\pi}{3} \left( T+\sqrt{T^2 + 3(\mu/4\pi)^2 + 6(\beta/4\pi)^2}  \right)\,,
\end{equation}
which is defined in \eqref{r0}.

Next, in the limit $\omega \rightarrow 0 $ with finite $\beta$
\begin{equation}
\sigma(\omega\rightarrow0) = K \tau  + \sigma_Q = 1+ \frac{\mu^2}{\beta^2} \,,
\end{equation}
 where \eqref{DC1} is used. 
Therefore, the relaxation time $\tau$ reads
\begin{align} \label{tauu}
\tau & = \frac{1+ \frac{\mu^2}{\beta^2} - \sigma_Q}{K}  \\  \label{tauu1}
&= \frac{1}{4\pi T} \cdot \frac{45 \tilde{\b}^4 +36 \tilde{\mu}^4 +2(1+\Delta)+6 \tilde{\b}^2(4+12 \tilde{\mu}^2
+3 \Delta )+3 \tilde{\mu}^2 (5+4\Delta)}{\tilde{\b}^2 (1+\Delta)(1+3 \tilde{\b}^2 +6 \tilde{\mu}^2 +\Delta)} \,,
\end{align}
where $\sigma_Q$ and $K$ is given in \eqref{Jo0} and 
\begin{equation}
\Delta \equiv \sqrt{1+3\tilde{\mu}^2 +6 \tilde{\b}^2}\,, \quad  \tilde{\mu} \equiv \frac{\mu}{4\pi T} \,, \quad \tilde{\beta} \equiv \frac{\beta}{4\pi T} \,.
\end{equation}
\begin{figure}[]
\centering
   {\includegraphics[width=8cm]{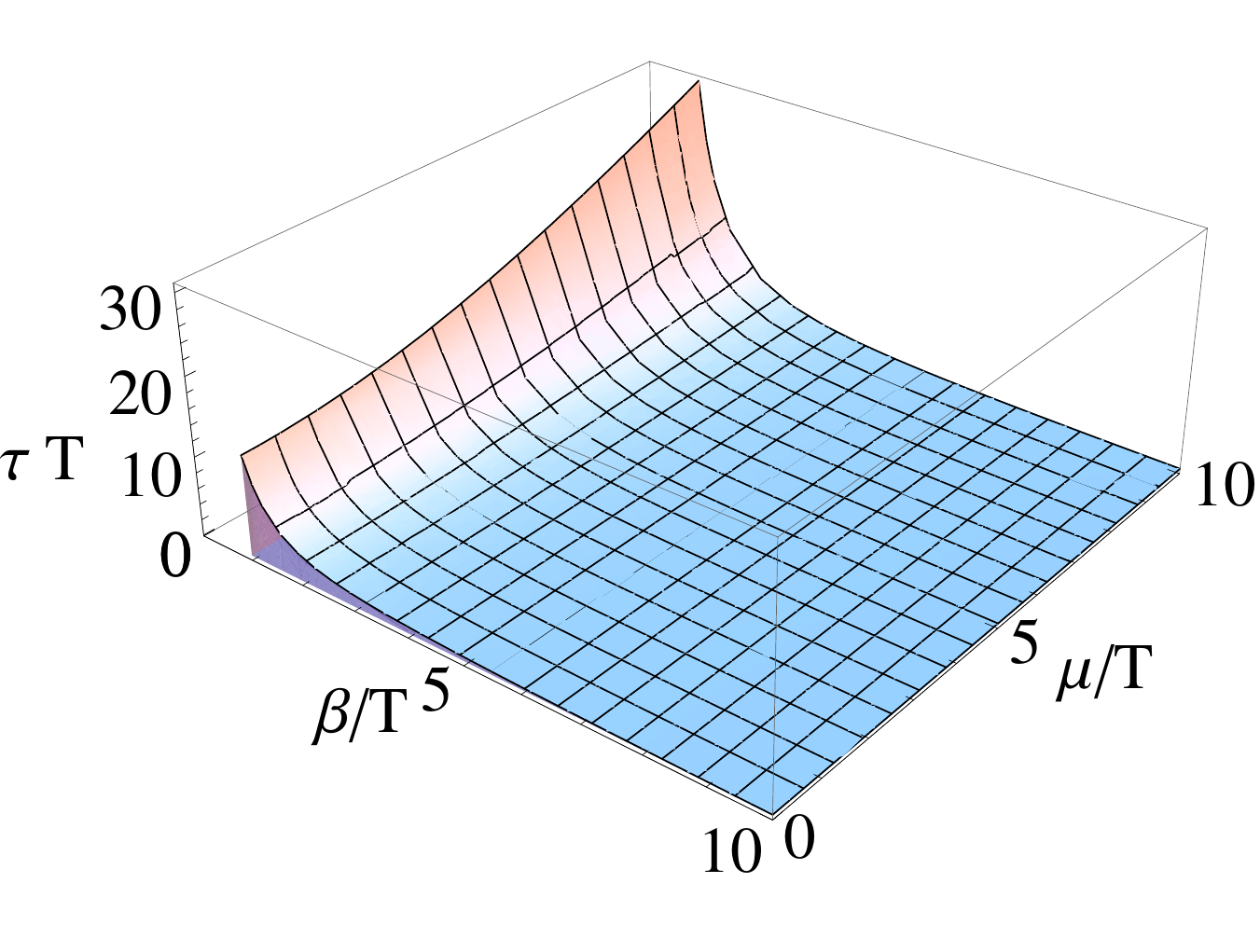} \label{}} 
 \caption{ Relaxation time $\tau$ at small $\omega$ as a function of $\mu/T$ and $\beta/T$. We do not plot the range $\beta/T < 1$ since  $\tau$ diverges quickly as $\beta$ goes to zero.  } \label{taufig}
\end{figure}
The expression \eqref{tauu1} is not very illuminating so we make a plot of the relaxation time as a function of 
$\mu/T$ and $\beta/T$ in Figure \ref{taufig}. The plot is meaningful only for the regime $\beta < \mu$ because it turns out that the Drude model \eqref{Drude00} works well for that regime(See Figure \ref{Drudes} and related discussion).  There is a tendency that a smaller $\beta$ and larger $\mu$ make the relaxation time longer, which is compatible with the interpretation of $\beta$ as an impurity effect.  For $T \ll \beta \ll \mu$, `clean limit'(small impurity) at low temperature, the relaxation time \eqref{tauu1} yields
\begin{equation}\label{tauu3}
\tau \approx 2\sqrt{3}\frac{\mu}{\beta^2} \,.
\end{equation}
\begin{figure}[]
\centering
  \subfigure[Re $\sigma$ at  ${\beta/\mu}=1/4$ ]
   {\includegraphics[width=4.8cm]{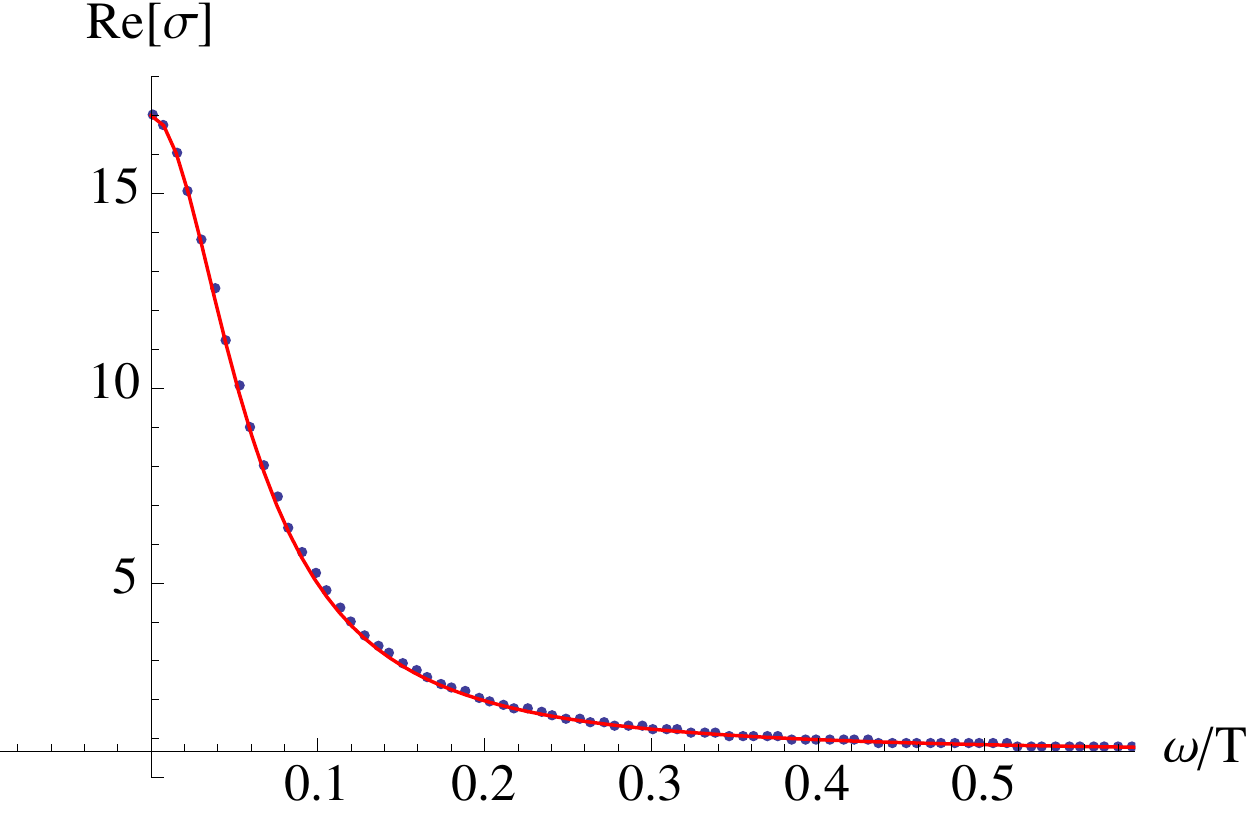} \label{}} 
 \subfigure[Re $\sigma$ at  ${\beta/\mu}=1/2$]
   {\includegraphics[width=4.8cm]{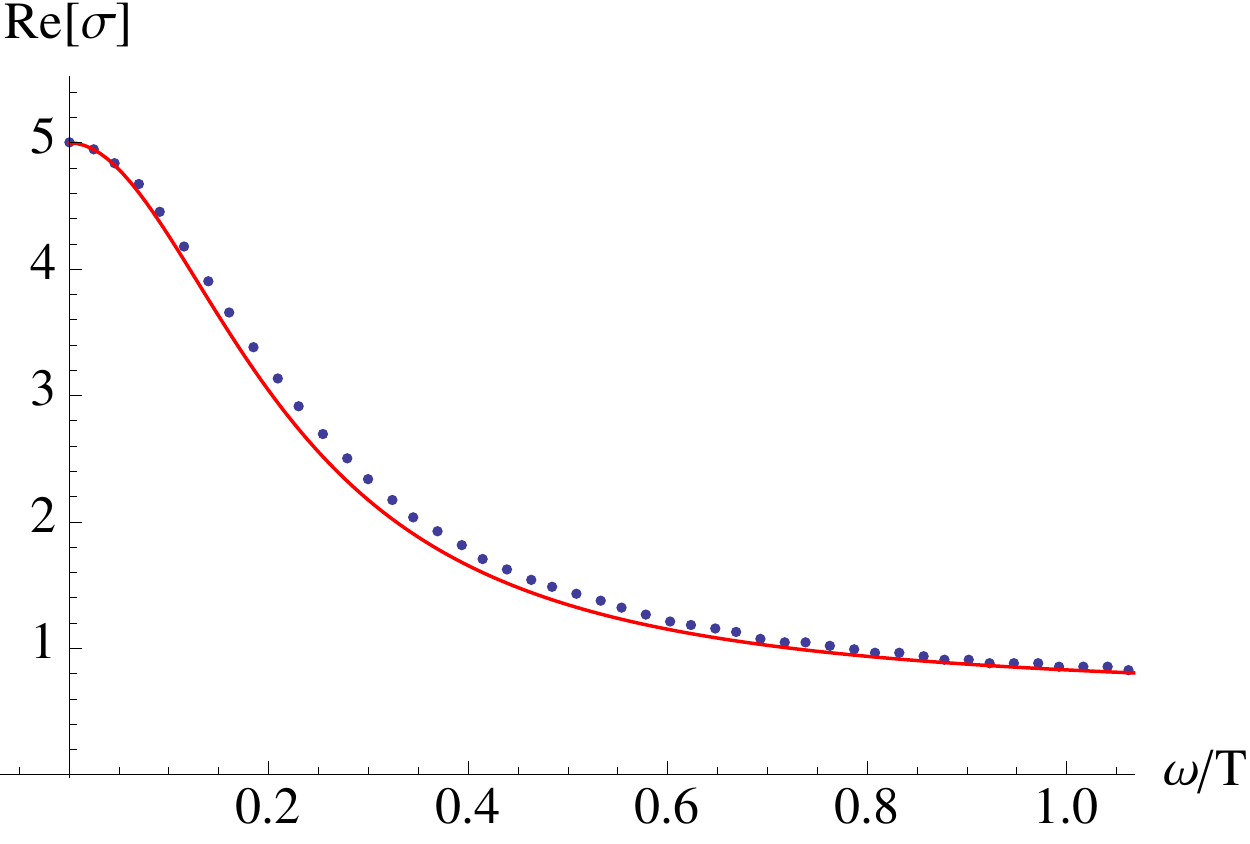} \label{} } 
  \subfigure[Re $\sigma$ at  ${\beta/\mu}=1$]  
      {\includegraphics[width=4.8cm]{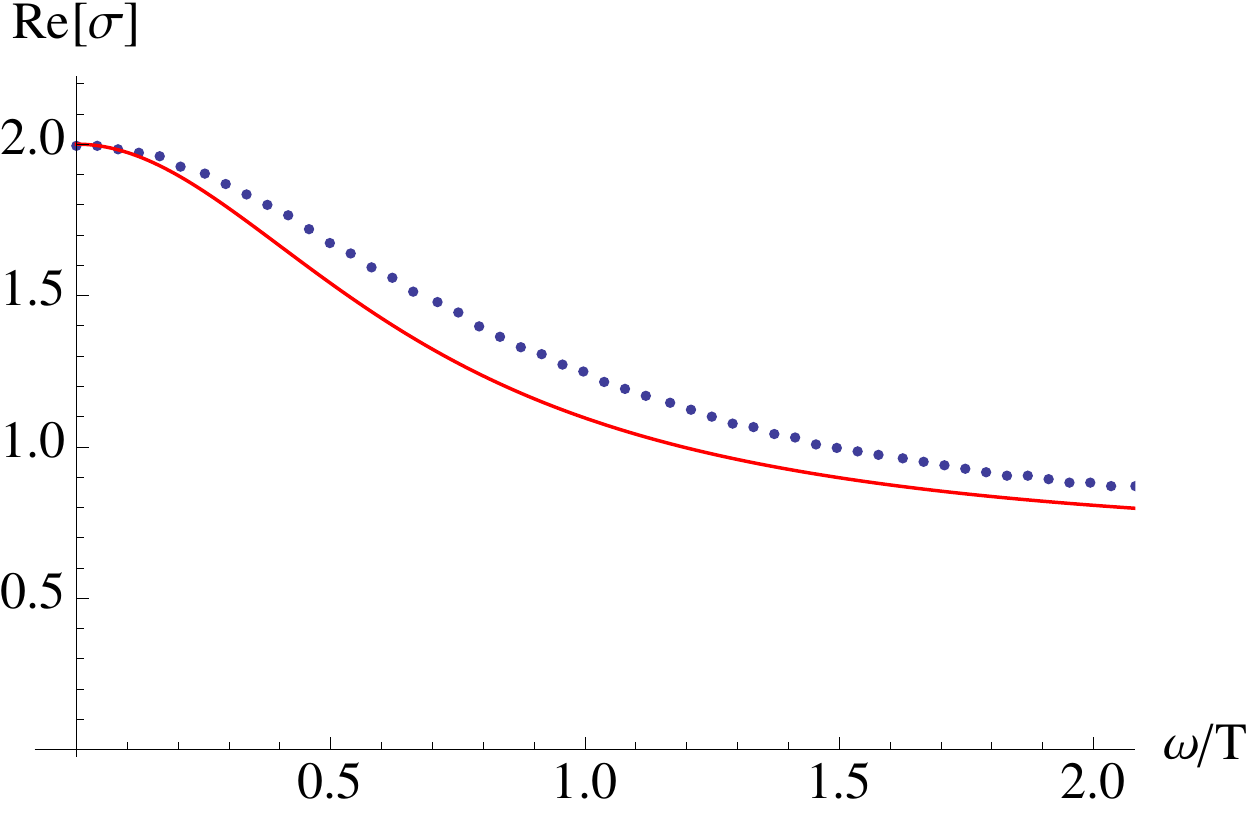} \label{}} 
     \subfigure[Im $\sigma$ at ${\beta/\mu}=1/4$]
   {\includegraphics[width=4.8cm]{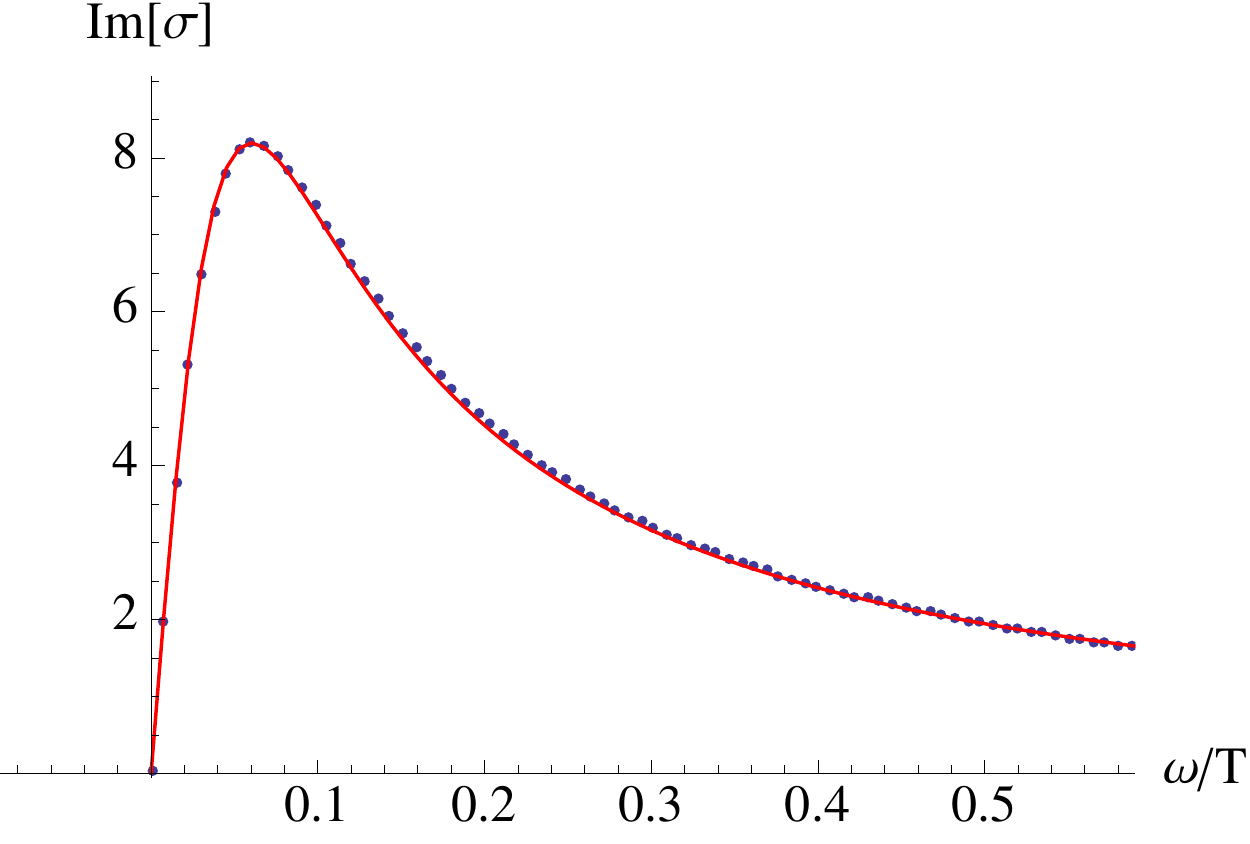} \label{} } 
    \subfigure[Im $\sigma$ at  ${\beta/\mu}=1/2$]
      {\includegraphics[width=4.8cm]{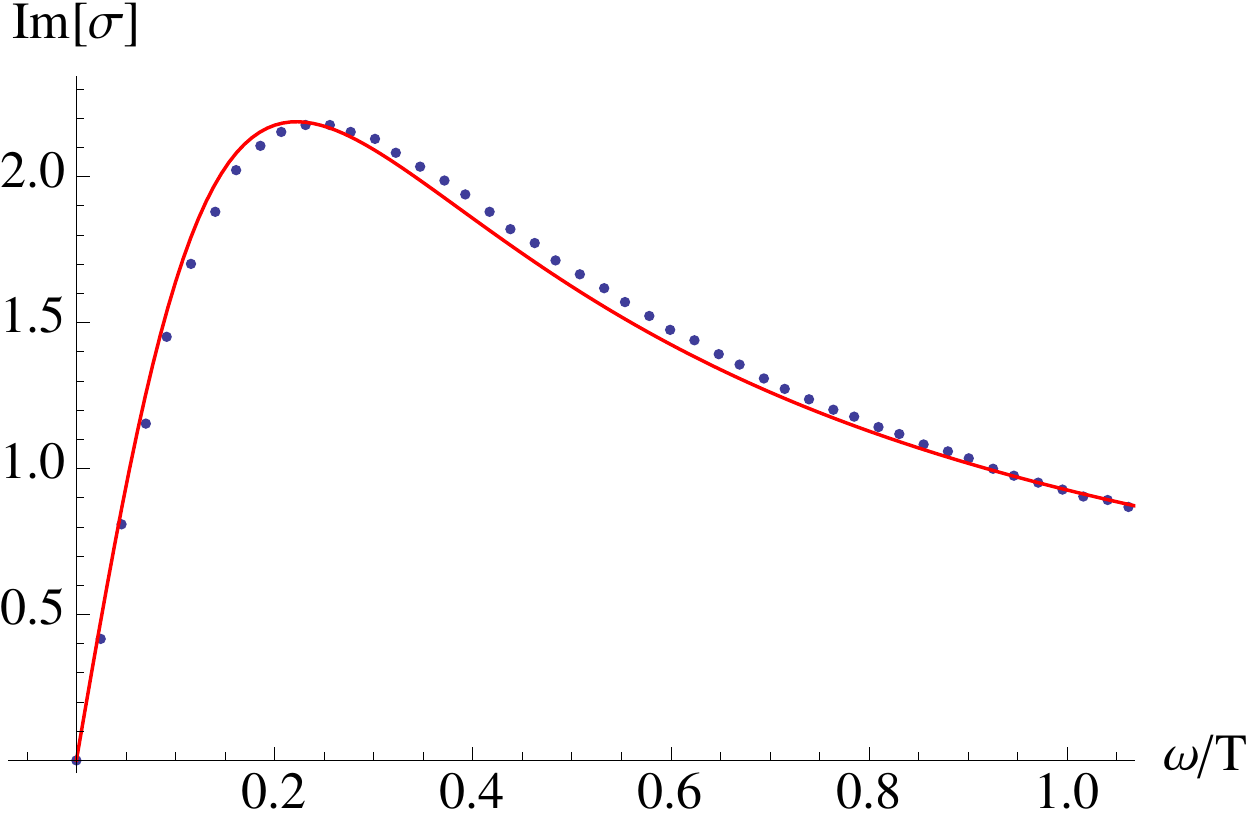} \label{}} 
     \subfigure[Im $\sigma$ at  ${\beta/\mu}=1$]
   {\includegraphics[width=4.8cm]{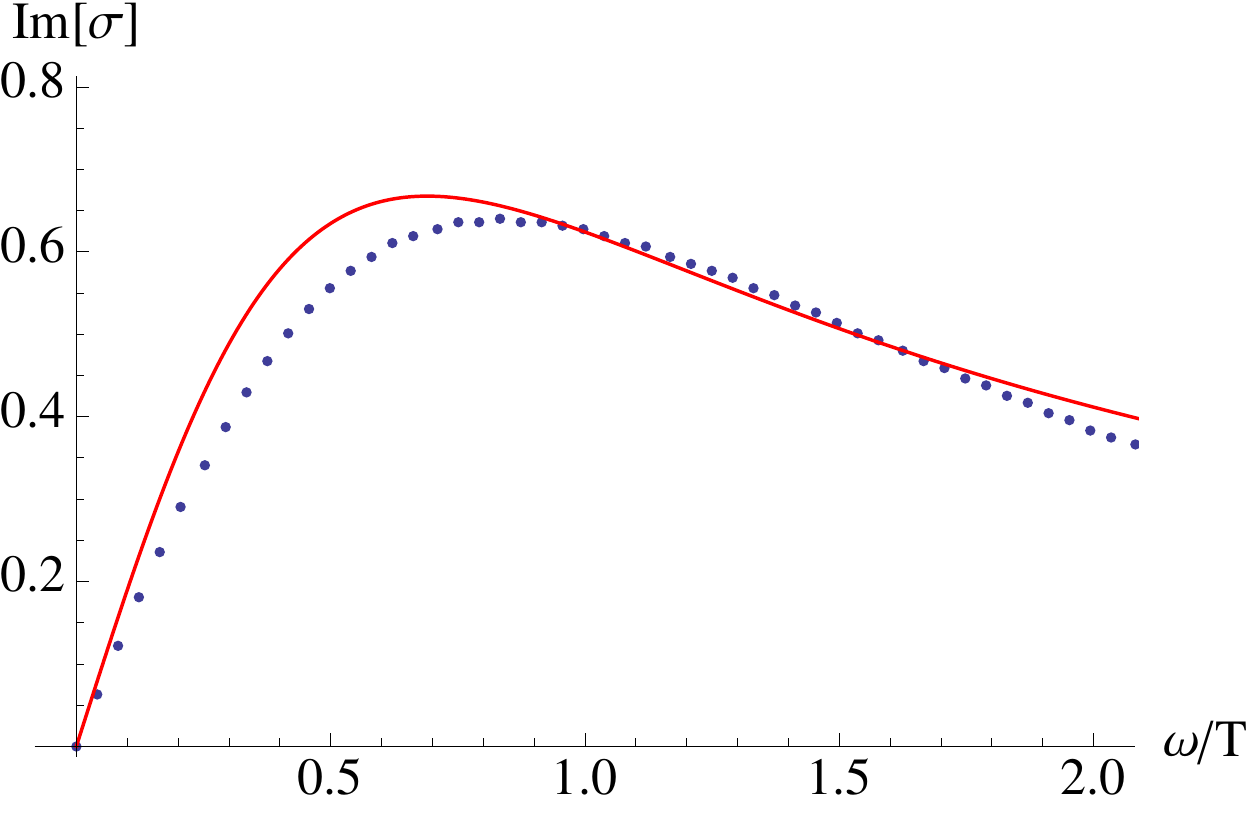} \label{} } 
 \caption{We compare numerical data(blue dotted lines) with a Drude model(red solid curves)\eqref{Drude00} of which parameters are fixed analytically in \eqref{Jo0} and \eqref{tauu}. ${\mu/T} = 4$.  When  ${\beta}/{\mu} \le 1/2 $ the numerical data agree well to the Drude model. The transition to incoherent metal is around $\beta/\mu \sim 1/2$.
           } \label{Drudes}
\end{figure}

To check the validity of our analytic expression of the Drude model \eqref{Drude00} with parameters \eqref{Jo0} and \eqref{tauu1}, we have made numerical plots for a wide range of parameters and compared with  \eqref{Drude00}. Figure \ref{Drudes}(a,d) and (c,f) are examples showing a good agreement of numerical data to \eqref{Drude00} and deviation from  \eqref{Drude00} respectively.\footnote{Similar plots were obtained independently by Blaise Gout{\'e}raux and Richard Davison and presented at the workshop, ``Holographic methods and application'', Iceland, August, 2014.}  Blue dotted lines are numerical data and red solid curves are  the analytic expression \eqref{Drude00}. In Figure \ref{Drudes} (c,f) if we find parameters $K, \sigma_Q, \tau$ by numerical fitting instead of using analytic expressions, the fitting curve is slightly improved, but it is still deviated from \eqref{Drude00}.
In these examples, when ${\beta}/{\mu} \le 1/2 $ the numerical data agree well to the Drude model.
In general, for small $\beta/\mu$, numerical data agrees well to a modified Drude model \eqref{Drude00} while for large $\beta/\mu$ the peak is not a Drude form. It is a concrete realisation of coherent/incoherent transition induced by impurity in a holographic model. In particular   if $\beta \ll \mu$(clean limit), the first term of \eqref{Drude00} is dominant and $\sigma_Q$ can be ignored. So \eqref{Drude00} is reduced to a standard Drude form. For example, with the parameters of Figure \ref{Drudes},  if $\beta/\mu < 1/6$,  numerical peaks are well fit to a standard Drude form.   If $\beta \gg \mu$(dirty limit) the peak is suppressed and becomes flat, approaching to $1$, which corresponds the limit $\mu \rightarrow 0$ (Figure \ref{sigmamu}(a)). 

Next, we want to investigate the scaling property in the intermediate frequency regime. 
In the range $T < \omega < \mu$, it was shown experimentally that certain high temperature superconductors in the normal phase exhibits scaling law
\begin{equation} \label{scaling00}
\sigma = \frac{B}{\omega^{\gamma}}e^{i\frac{\pi}{2} \gamma} \sim \left( \frac{i}{\omega}  \right)^{\gamma} \,,
\end{equation}
where $\gamma \approx 2/3$ and $B$ is constant~\cite{Marel:2003aa}.
This scaling has been discussed also in holographic models with momentum dissipation. In models studied in\cite{Horowitz:2012ky,Horowitz:2012gs,Ling:2013nxa}  modified scalings \eqref{modscaling} have been reported while in \cite{Donos:2013eha,Taylor:2014tka,Donos:2014yya} no scaling law have been observed. With our model we have analysed  several cases for a wide range of parameters to search a scaling behaviour. However it seems that there is no robust scaling law.

\begin{figure}[]
\centering
  \subfigure[$\beta/r_0=0.1$ ]
   {\includegraphics[width=4.6cm]{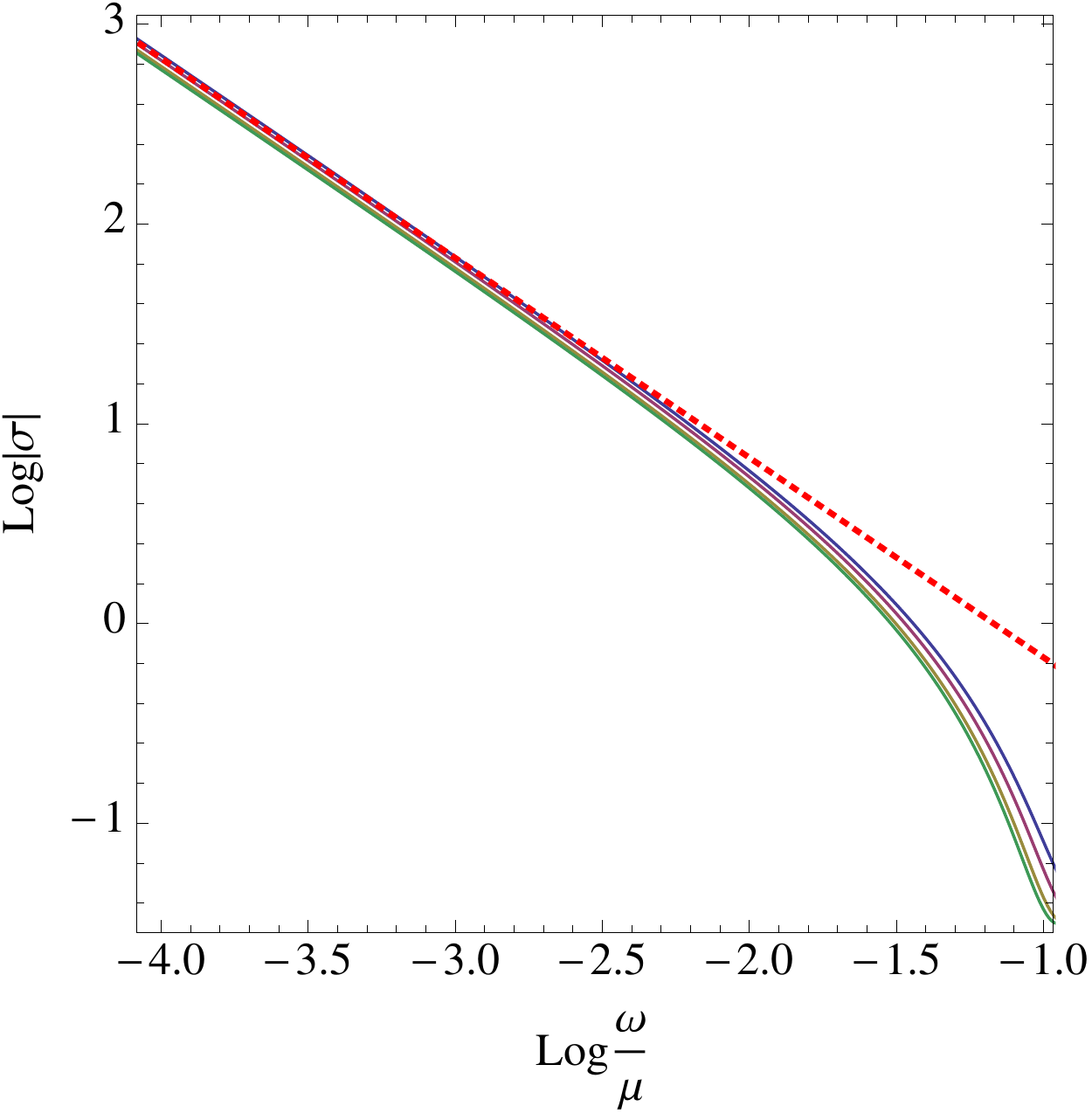} \label{}} 
     \subfigure[$\beta/r_0=1$]
   {\includegraphics[width=4.6cm]{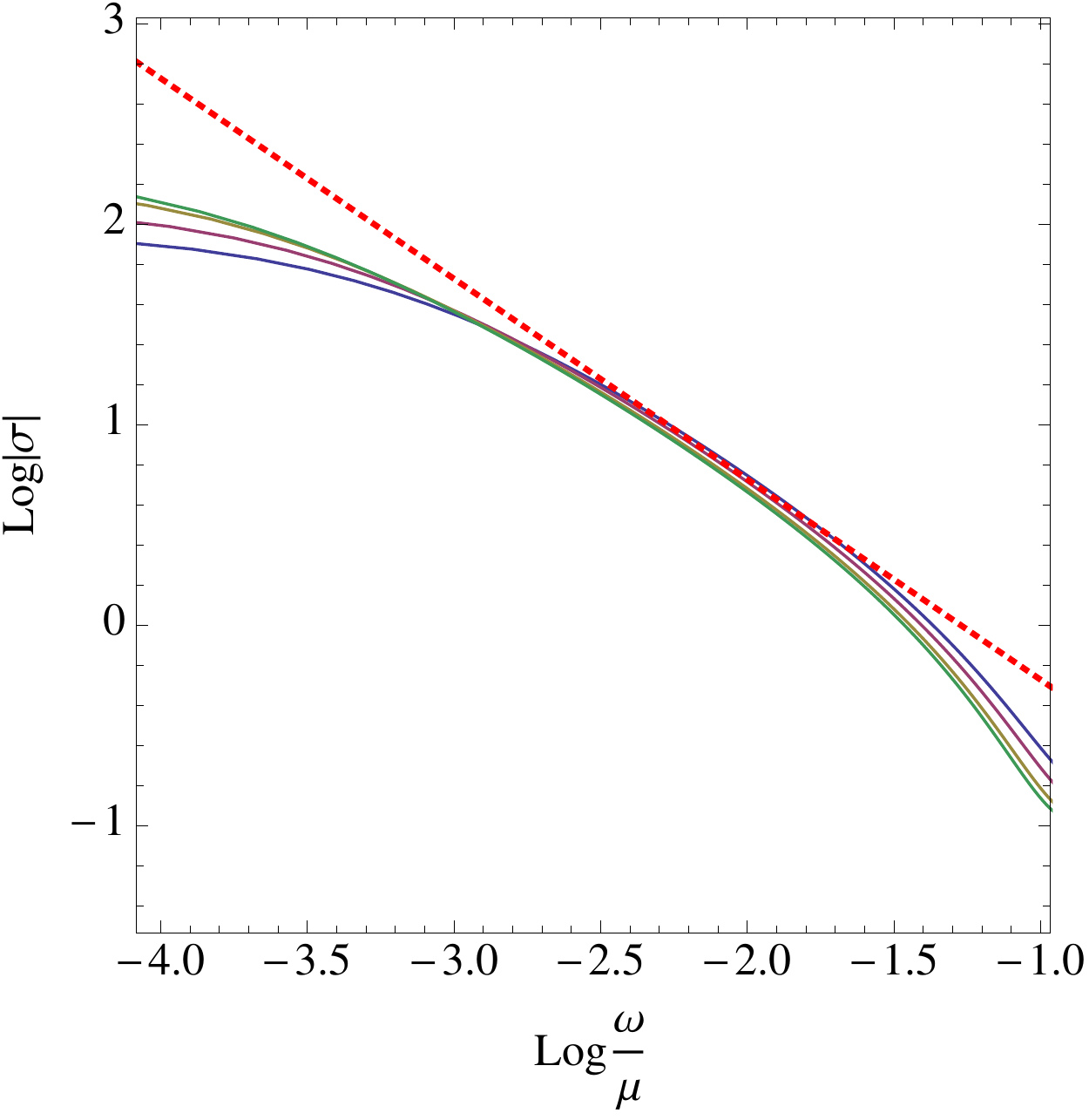} \label{} } 
        \subfigure[$\beta/r_0=1.5$]
   {\includegraphics[width=4.6cm]{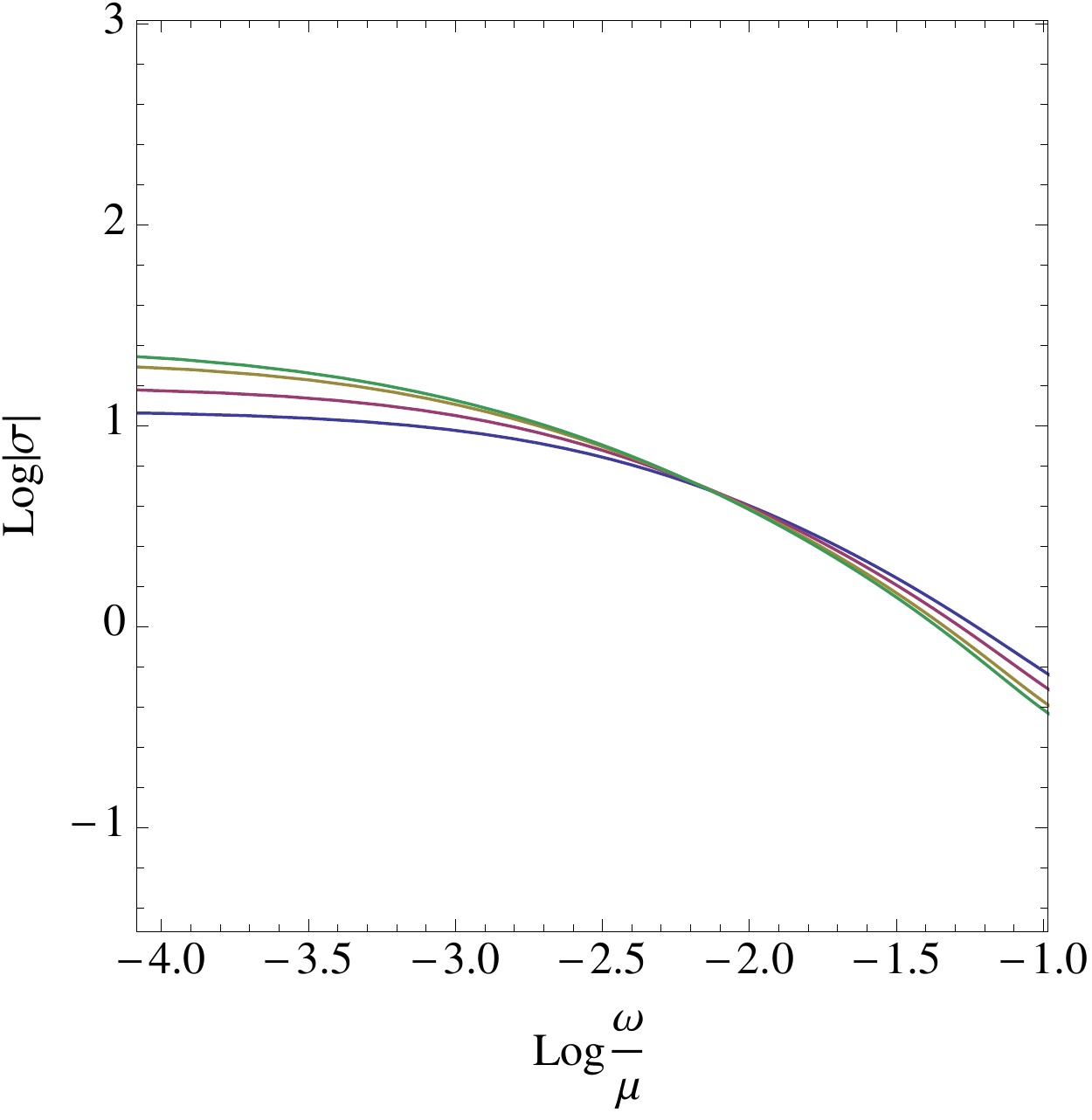} \label{} } 
 \caption{Search for scaling behaviour at intermediate $\omega$($T < \omega< \mu$).   Four curves are for $T/\mu = 0.005,0.01,0.02,0.03$.  The slope of red dotted lines in (a) and (b) is $-1$, which is a signal of Drude model at small $\omega$. } \label{scalings}
\end{figure}
\begin{figure}[]
\centering
  \subfigure[$\gamma \approx 0.24$ ]
   {\includegraphics[width=6cm]{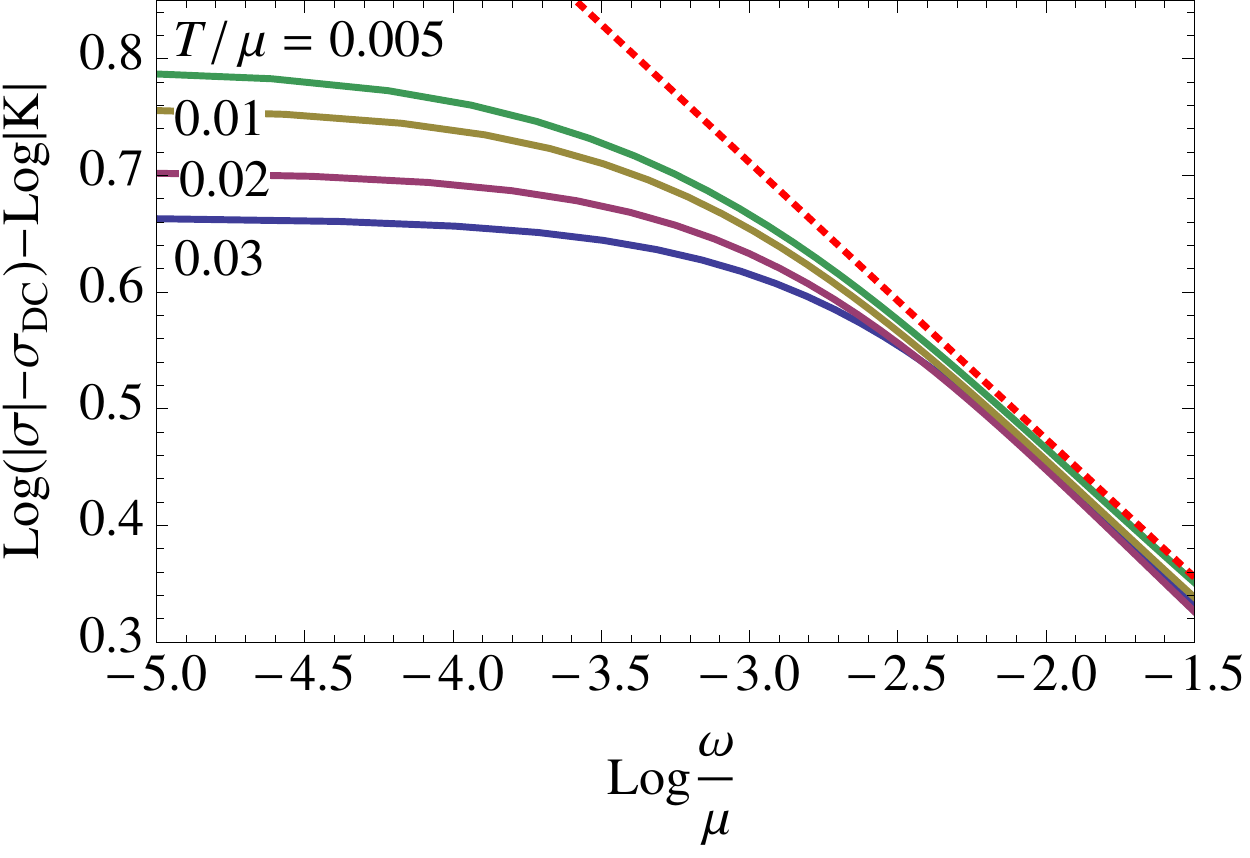} \label{}}  \hspace{5mm}
     \subfigure[$\tilde{\gamma} \approx 2 \gamma$]
   {\includegraphics[width=6cm]{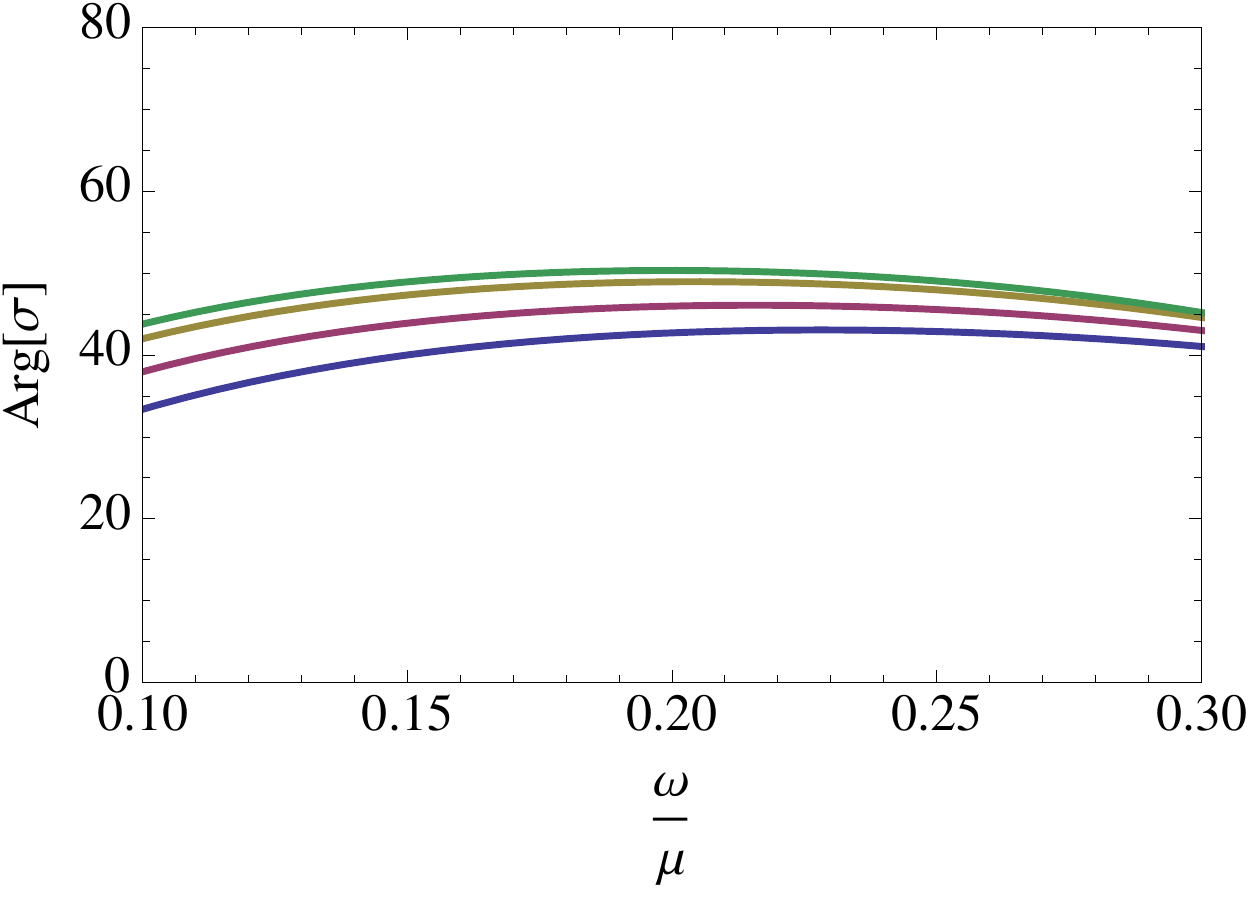} \label{} } 
 \caption{Approximate scaling behaviour ($\beta/r_0=1.5$, Figure \ref{scalings}(c)). This is not a precise and robust scaling law under change of parameters. These plots are presented to show how the constant shift $C$ in \eqref{modscaling} can improve the scaling behaviour.} \label{scalingsk}
\end{figure}

For completeness, in Figure \ref{scalings}, we present a typical behaviour of $\abs{\sigma}$ at intermediate $\omega$ regime as $\beta$ is changed. 
When $\beta$ is small(Figure \ref{scalings}(a))  there is  a robust scaling  $\abs{\sigma} \sim \omega^{-1}$ for small $\omega$. As a guide we showed the red dotted lines in Figure \ref{scalings} (a) and (b) of which slopes are  $-1$.  This scaling can be understood as a tail of Drude form, because in this regime the Drude form is dominant as shown in Figure  \ref{sigmabeta} and \ref{Drudes}.  
As $\beta$ increases the scaling of Drude tail becomes weaken (Figure \ref{scalings}(b)) and disappears at bigger $\omega$(Figure \ref{scalings}(c))). We do not see a scaling behaviour of  the form \eqref{scaling00}.

Now we want to investigate if there is a modified scaling law motivated by previous holographic models~\cite{Horowitz:2012ky,Horowitz:2012gs,Ling:2013nxa,Davison:2013jba}.
\begin{equation} \label{modscaling}
\sigma =\left( \frac{B}{\omega^{\gamma}} + C\right)e^{i\frac{\pi}{2} \tilde{\gamma}} \,,
\end{equation}
where $B$ and $C$ are constants and $\tilde{\gamma}$ may be different from $\gamma$. We find that the Figure \ref{scalings}(c) can be approximately fitted by a modified scaling law, with $\gamma \approx 0.24 $
\begin{equation}
\sigma =\left( \frac{K}{(\omega/\mu)^{\gamma}} + \sigma_{DC}\right)e^{i \pi {\gamma}}\,, 
\end{equation}
which is shown in Figure \ref{scalingsk}. 
Interestingly, in this case, the constants $B$ and $C$ in \eqref{modscaling} are fixed by analytic $K$ and $\sigma_{DC}$, while in other previous studies, they are numerically determined.  However, 
this approximate scaling behaviour is not precise\footnote{A good way to check the constancy of $\gamma$ is to compute  $1+\omega \abs{\sigma}''/\abs{\sigma}'$ \cite{Donos:2013eha,Donos:2014yya}.} and robust under change of parameters.   We present Figure \ref{scalingsk} to show how the constant shift $C$ in \eqref{modscaling} can improve the scaling behaviour of Figure \ref{scalings}(c) even though it is not an evidence of a scaling behaviour.  After numerical analysis with a wide variety of parameters and cases we do not see a scaling behaviour of  the form \eqref{modscaling}, which agrees to the conclusion in \cite{Taylor:2014tka}.

\subsection{Thermoelectric and thermal conductivity}

Finally we plot the thermoelectric($\alpha$) and thermal($\bar{\kappa}$) conductivity in Figure \ref{thermo}. Qualitative feature is similar to electric conductivity. The red dots at $\omega = 0$ is the DC conductivities analytically computed in \cite{Donos:2014cya}
\begin{equation}
 \alpha = \frac{4\pi\mu}{\beta^2} r_0 \,, \quad \frac{\bar{\kappa}}{T} = \frac{(4\pi)^2}{\beta^2}   r_0^2 \,,
\end{equation}
At large $\omega$ it can be shown from Ward identity \cite{future1}
\begin{equation}
\alpha \rightarrow  -\frac{\mu }{T}  \,, \qquad \frac{\bar{\kappa}}{T} \rightarrow   \frac{\mu^2+ \beta^2}{T^2}
\end{equation}
 Numerical plots in Figure \ref{thermo} shows a good agreement to both limits($\omega \rightarrow 0$ and $\omega \rightarrow \infty$).

\begin{figure}[]
\centering
  \subfigure[Re ${\alpha}$ ]
   {\includegraphics[width=6cm]{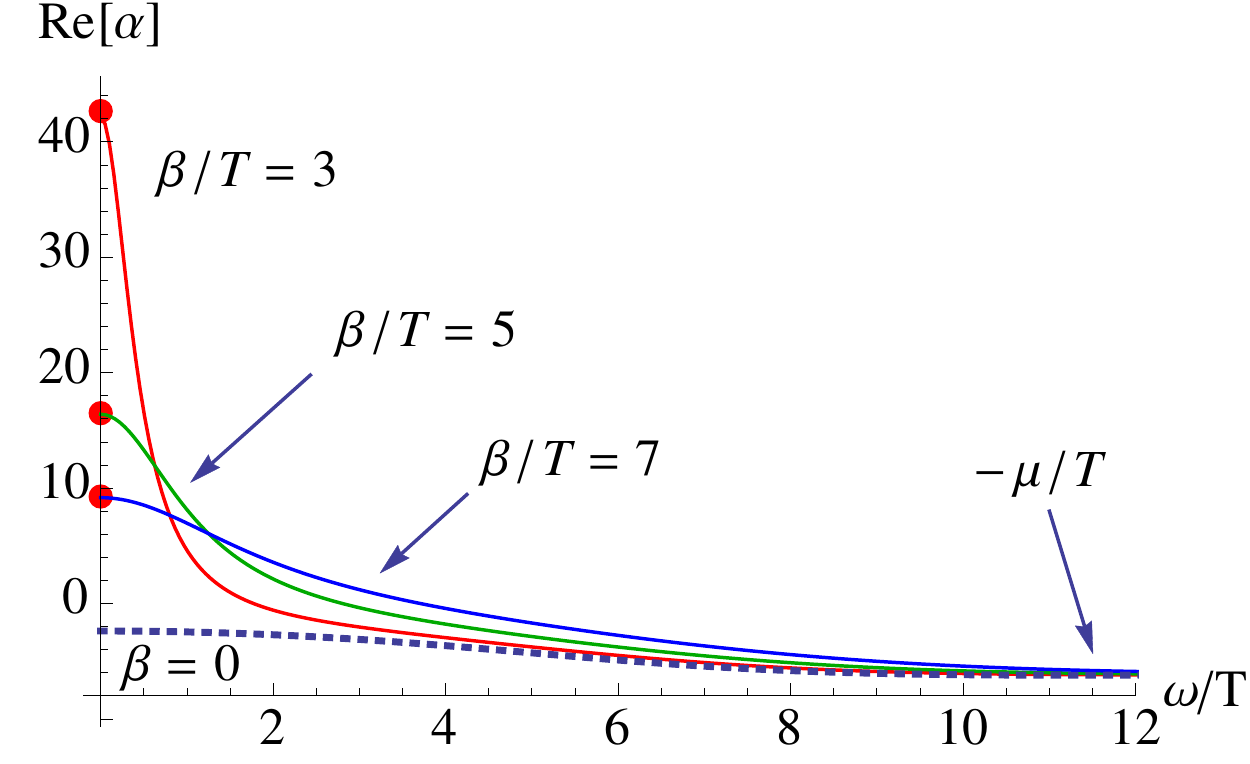} \label{}}  \hspace{5mm}
     \subfigure[Re $\bar{\kappa}/T$]
   {\includegraphics[width=6cm]{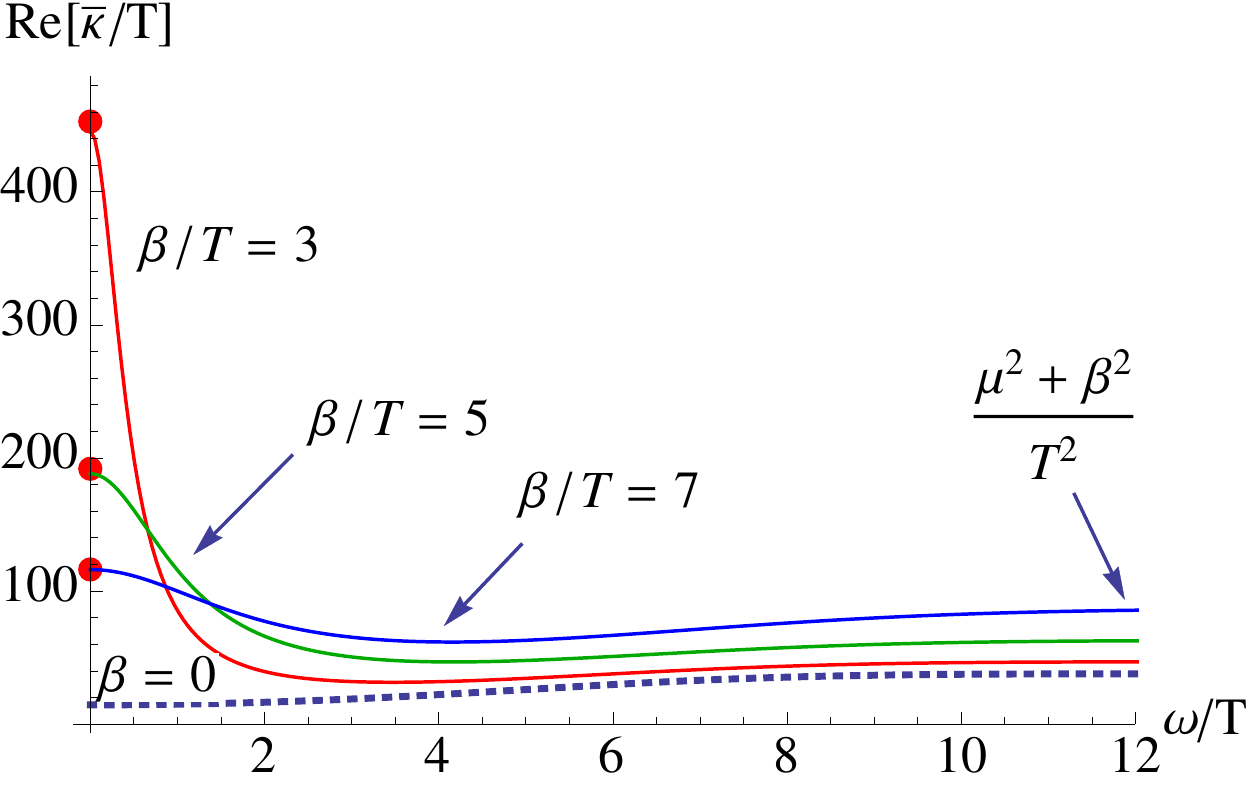} \label{} } 
 \caption{ Thermal and thermoelectric conductivity with momentum relaxation at fixed $\mu/T = 6$. As we increase $\beta$, the Drude peak disappears and the transition to incoherent metal is manifest.
           } \label{thermo}
\end{figure}

In order to discuss the Wiedemann-Franz law, we compute the ratio of the DC thermal conductivity to the DC electric conductivity as follows
\begin{equation}
\bar{L} \equiv \frac{\bar{\kappa}}{\sigma T} = \frac{1}{\tilde{\mu}^2 + \tilde{\beta}^2} \frac{r_0^2}{T^2} = \frac{4\pi^2 \left(1+\sqrt{1+3(2\tilde{\beta}^2 + \tilde{\mu}^2)}\right)^2}{9(\tilde{\beta}^2+\tilde{\mu}^2)}\approx \frac{4\pi^2}{3} \frac{ 2 \beta^2 +  \mu^2}{  \beta^2 +  \mu^2 }~~, 
\end{equation}
where we took low temperature limit in the last expression since the Wiedemann-Franz law is supposed to be valid at low temperature.  In two extreme limits, in the clean($\beta \ll \mu$) and dirty($\beta \gg \mu$) limit, the ratio becomes constant
\begin{align}
\bar L = \left\{ ~~  \begin{array}{c}  \frac{4\pi^2}{3} ~~~~( \mu  \gg   \beta)   \\ \frac{8\pi^2}{3} ~~~~( \beta \gg  \mu)\end{array}   \right. \,.
\end{align}
but the numerical values are different from the Fermi-liquid case, as expected in a non-Fermi liquid, see e.g. \cite{Kim:2009aa}. 

At small frequencies, like electric conductivity, thermoelectric and thermal conductivities also have a modified Drude peak similar to 
\eqref{Drude00} for $\beta < \mu$, 
\begin{equation} \label{Drude000}
 \alpha(\omega) = \frac{A_\alpha \tau_\alpha}{1 - i \omega \tau_\alpha} + B_\alpha \,, \qquad  \frac{\bar{\kappa}(\omega)}{T} = \frac{A_{\bar{\kappa}} \tau_{\bar{\kappa}}}{1 - i \omega \tau_{\bar{\kappa}}} + B_{\bar{\kappa}}
\end{equation}
while the peak is non-Drude for $\beta > \mu$. Like $K, \sigma_Q, \tau$ in \eqref{Drude00}, $A_\alpha, B_\alpha,\tau_\alpha, A_{\bar{\kappa}}, B_{\bar{\kappa}}, \tau_{\bar{\kappa}} $ may be obtained analytically by using the hydrodynamics results in \cite{Ge:2010yc,Matsuo:2009yu,Ge:2008ak}. Figure \ref{Drudes1} shows an excellent agreement of numerical data to \eqref{Drude000}, where the blue dots are numerical values and the red solid curve is a fitting to \eqref{Drude000}. The relaxation times $\tau, \tau_\alpha$, and $\tau_{\bar{\kappa}}$ are defined by comparing with Drude model, which works only when $\beta \le \mu$.  In this  regime, 
we find that three relaxation time are almost the same.  This is because  in all three cases the Drude peaks  are due to the momentum relaxation. So we have only one $\tau$.\footnote{ We thanks the referee for pointing this out.}
At intermediate frequencies, we do not see any scaling law unlike \cite{Horowitz:2012gs}.

\begin{figure}[]
\centering
  \subfigure[Re ${\alpha}$ ]
   {\includegraphics[width=6cm]{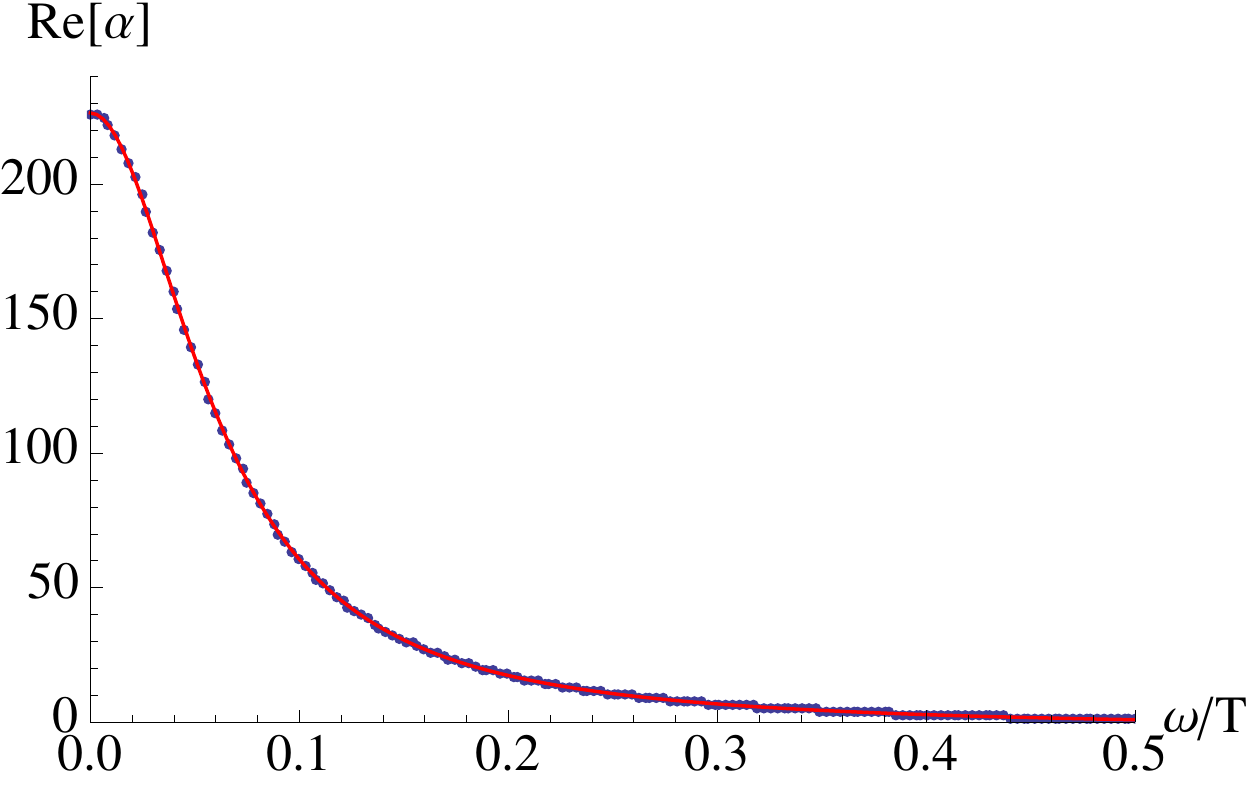} \label{}}  \hspace{5mm}
     \subfigure[Im $\alpha$]
   {\includegraphics[width=6cm]{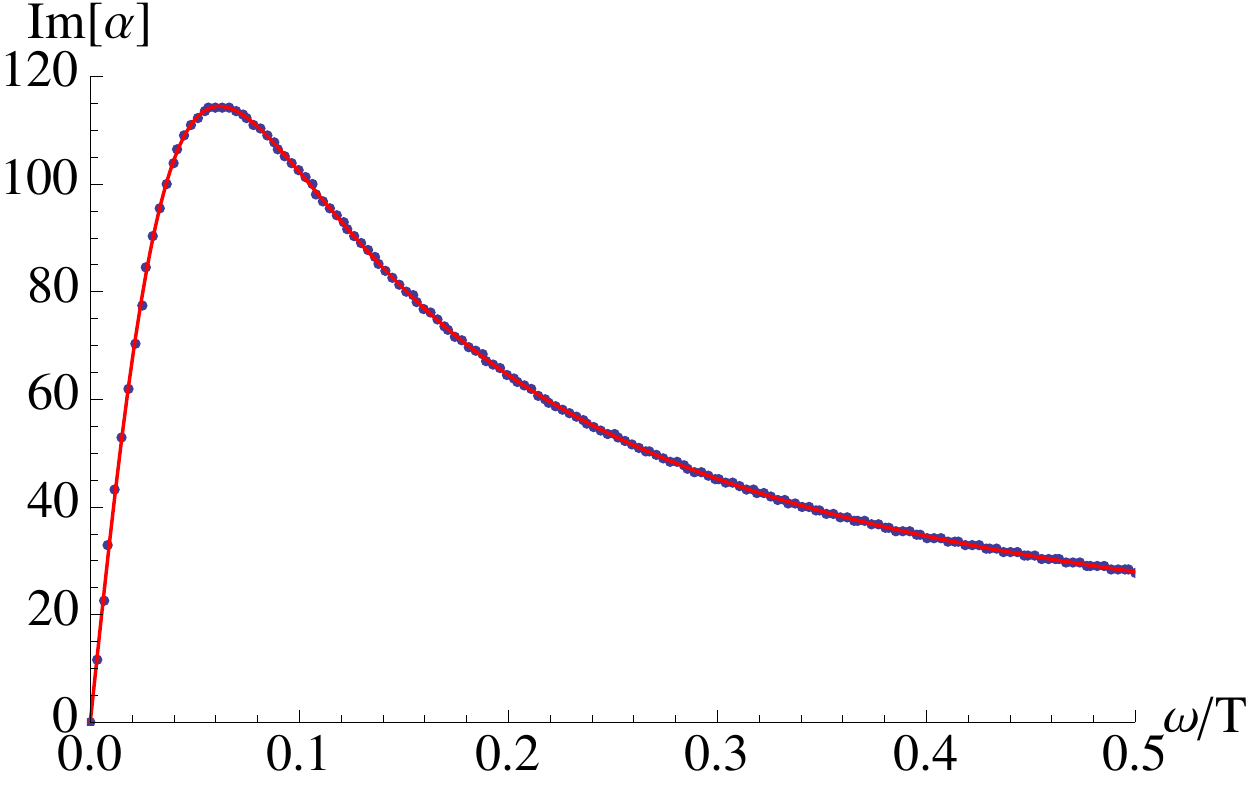} \label{} } 
     \subfigure[Re ${\bar{\kappa}/T}$ ]
   {\includegraphics[width=6cm]{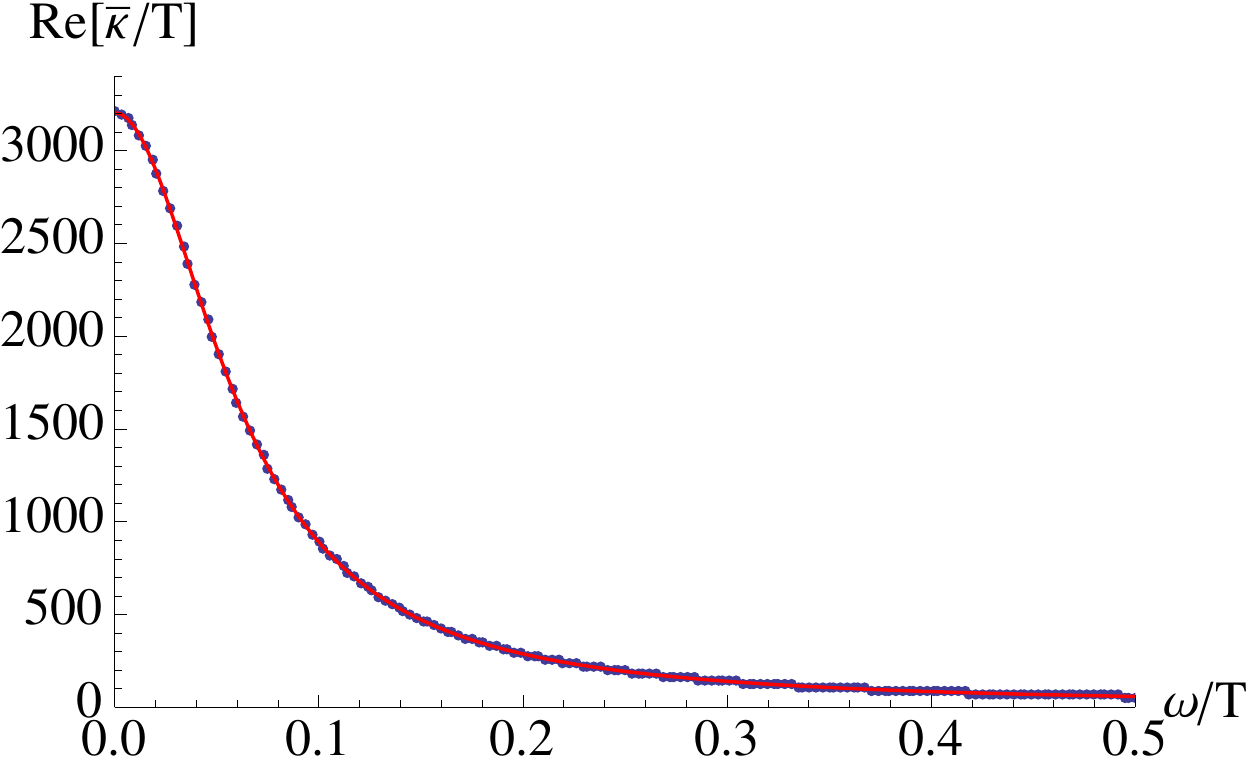} \label{}}  \hspace{5mm}
     \subfigure[Im $\bar{\kappa}/T$]
   {\includegraphics[width=6cm]{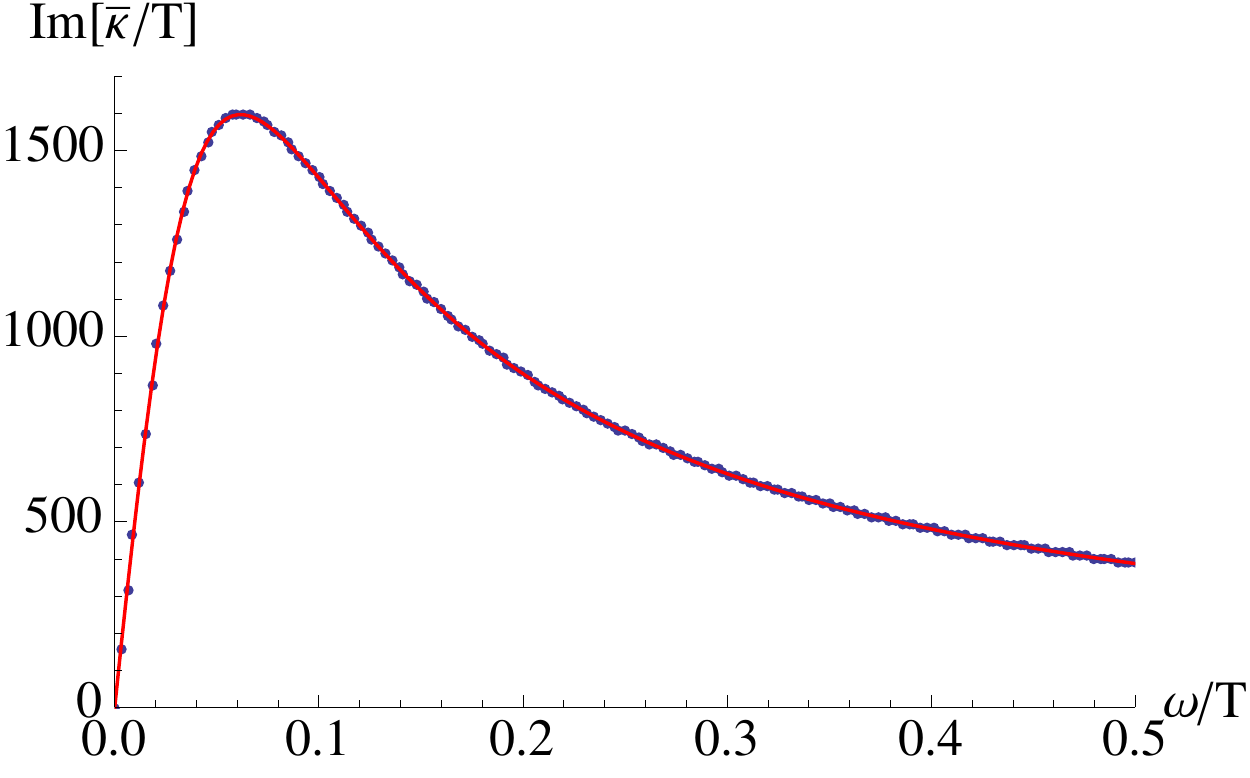} \label{} } 
 \caption{ Drude peaks of thermoelectric($\alpha$) and thermal($\kappa$) conductivity with the same parameters as  Figure \ref{Drudes}(a): $\mu/T = 4$ and $\beta/\mu = 1/4$. Blue dots are numerical data and the red solid lines are  fitting curves of the from \eqref{Drude000}.
           } \label{Drudes1}
\end{figure}

\section{Conclusions} \label{Conc}

In this paper,  we study three conductivities(electric($\sigma$), thermoelectric($\alpha$), and thermal($\bar{\kappa}$) conductivities) in a holographic model of momentum relaxation~\cite{Andrade:2013gsa}. The model is based on the 3+1 dimensional Einstein-Maxwell-scalar action.  Momentum is dissipated due to massless scalar fields linear to every spatial coordinate. 
\begin{equation} \label{con1}
\psi_1 =  \beta  x\,,  \qquad \psi_2 = \beta y 
\end{equation}
where $x, y$ are spatial coordinates in field theory. It turns out the $\beta$ plays a role of impurity.  There are two more free parameters in the model: temperature($T$) and chemical potential($\mu$). 
The background bulk metric and gauge fields compatible with \eqref{con1} are given analytically. They depend on only holographic direction because the scalar field enters the stress tensor through the derivative($\partial_M\psi_i$).  The fluctuation fields(metric, gauge, and scalar fields) relevant for three conductivities can be chosen to be functions of only the holographic direction, so the computations can be done by coupled ODEs rather than PDEs. 

Our numerical method reproduces the previous AC conductivities($\sigma,\alpha,\bar{\kappa}$) at $\beta=0$ \cite{Hartnoll:2009sz} and matches the analytic values at $\omega =0 $ \cite{Andrade:2013gsa, Donos:2014cya} and at $\omega \rightarrow \infty$ \cite{future1} at finite $\beta$. At $\omega=0$, in both the clean($\beta \ll \mu$) and dirty($\beta \gg \mu$) limit  the ratio $\bar{\kappa}/\sigma T$  approaches temperature independent constants, but the numerical values are different from the Fermi-liquid case. 
 We presented a concrete realisation of coherent/incoherent transition induced by impurity in a holographic model.
At low frequencies, if $\beta < \mu$ (coherent metal phase) all three conductivities show a modified Drude peak.  For example, for electric conductivity,  
\begin{equation} \label{DrudeIntro1}
\sigma(\omega) = \frac{K \tau}{1 - i \omega \tau} + \sigma_Q \,,
\end{equation}
where $\sigma_Q$ denote a contribution from pair production. We have obtained the analytic formula for $K, \tau$ \eqref{Jo0} and $\sigma_Q$ \eqref{tauu}. The same Drude from is found for $\alpha$ and $\bar{\kappa}$, but with different parameter values. For example, the relaxation times are different for three conductivities in general.  
In the clean limit $\beta \ll \mu$, $\sigma_Q$ can be ignored and \eqref{DrudeIntro1} becomes a standard Drude from with $\tau \approx  2\sqrt{3} \frac{\mu}{\beta^2}$ \eqref{tauu3}. For $\beta > \mu$ (incoherent metal phase) the peak is not Drude-like. In the dirty limit $\beta \gg \mu$, the peak disappears and becomes flat, approaching  $1$ for all $\omega$, which amounts to the limit $\mu \rightarrow 0$ (Figure \ref{sigmamu}(a)). In all cases, a sum rule is satisfied. i.e. the area of peaks due to momentum relaxation($\beta \ne 0$) is always the same as the area of the delta function at $\beta=0$. There is a finite plateaux region at large $\omega$ in AC conductivity  due to the massless nature of the charge carrier as mentioned below \eqref{Ttt}. If we can use massive one the constant plateaux will disappear.   

At intermediate frequencies, $T < \omega < \mu$,  we have tried to find scaling laws such as
\begin{equation}
\sigma = \frac{B}{\omega^{\gamma}}e^{i\frac{\pi}{2} \gamma} \sim \left( \frac{i}{\omega}  \right)^{\gamma} \,, \qquad \sigma =\left( \frac{B}{\omega^{\gamma}} + C\right)e^{i\frac{\pi}{2} {\tilde{\gamma}}}\,,
\end{equation}
where $\gamma, \tilde{\gamma}, B$ and $C$ are constant.
These scalings are motivated by experiments \cite{Marel:2003aa} and some holographic models~\cite{Horowitz:2012ky,Horowitz:2012gs}.  but we find no robust scaling law, which agrees to the conclusion in \cite{Donos:2013eha,Taylor:2014tka,Donos:2014yya}. In \cite{Bhattacharya:2014dea} a mechanism to engineer scaling laws was provided, where translation symmetry is not broken. It would be interesting to generalize it to our case.

Without momentum dissipation, the three conductivities($\sigma,\alpha,\bar{\kappa}$) are simply related by Ward identities, and once electric conductivity is given the other two are algebraically determined~\cite{Hartnoll:2009sz,Herzog:2009xv}.  In our model the relationship between them are more complicated, involving the background scalar fields. It will be interesting to understand how their relationship are modified by $\beta$.

We introduced a general numerical method to compute the holographic retarded Green's functions when many fields are coupled. This method,  used to compute three conductivities in this paper, can be applied also to other models and problems such as \cite{Donos:2013eha, Donos:2014uba, Gouteraux:2014hca, Donos:2014oha,Taylor:2014tka}.
It would be interesting to extend our analysis to dyonic black holes and holographic superconductors~\cite{future1}. 
It would be also interesting to study the models based on other free massless form fields introduced in \cite{Bardoux:2012aw}, which may be used to engineer certain desired properties of condensed matter systems.

\acknowledgments
We would like to thank Yunkyu Bang, Richard Davison, Aristomenis Donos, Jerome Gauntlett, Xian-Hui Ge, Blaise Gouteraux, 
  Takaaki Ishii, Yan Liu, Ya-Wen Sun, and Marika Taylor for valuable discussions and correspondence.
The work of KYK and KKK was supported by Basic Science Research Program through the National Research Foundation of Korea(NRF) funded by the Ministry of Science, ICT \& Future Planning(NRF-2014R1A1A1003220). 
The work of SS and YS was supported by Mid-career Researcher Program through the National Research Foundation of Korea (NRF) grant No. NRF-2013R1A2A2A05004846. YS was also supported in part by Basic Science Research Program through NRF grant No. NRF-2012R1A1A2040881.
We acknowledge the hospitality at APCTP(``Aspects of Holography'', Jul. 2014) and Orthodox Academy of Crete(``Quantum field theory, string theory and condensed matter physics'', Sep. 2014), and at CERN (``Numerical holography'', Dec. 2014), 
where part of this work was done.

\bibliographystyle{JHEP}
\bibliography{KyKimRefs}

\end{document}